\documentclass[twocolumn]{aastex7}
\usepackage{xcolor}
\usepackage{xspace}
\usepackage{amsmath}
\usepackage{CJKutf8}
\usepackage{threeparttable}
\usepackage{tabularx}
\usepackage{makecell}
\usepackage{multirow}
\usepackage{orcidlink}
\usepackage{natbib}
\usepackage{enumerate}
\usepackage{amssymb}
\usepackage{graphicx}
\graphicspath{{Figures/}}

\begin{document}
\newcommand{\Exp}[1]{\exp{\left\{#1\right\}}}
\newcommand{\Int}{\int\limits_{-\pi}^{\pi}}
\newcommand{\hb}[1]{\hat{\mathbf{#1}}}
\newcommand{\F}{\mathcal{F}}
\newcommand{\EY}[1]{\color{orange}{#1} \color{black}}
\newcommand{\YS}[1]{\color{red}{#1} \color{black}}
\renewcommand{\Re}{\mathrm{Re}}
\renewcommand{\Im}{\mathrm{Im}}

\begin{CJK*}{UTF8}{gbsn}
\title{An Analytical Model for the Eccentricity Cascade: Hot Jupiter Formation via S-type Instability}
\author{Eritas Yang (杨晴) \orcidlink{0009-0005-2641-1531}}
\affiliation{Department of Astrophysical Sciences, Princeton University, 4 Ivy Lane, Princeton, NJ 08540, USA}
\email{eritas.yang@princeton.edu}
\author{Yubo Su (苏宇博) \orcidlink{0000-0001-8283-3425}}
\affiliation{Department of Astrophysical Sciences, Princeton University, 4 Ivy Lane, Princeton, NJ 08540, USA}
\affiliation{Canadian Institute for Theoretical Astrophysics, 60 St. George Street, Toronto, ON M5S 3H8, Canada}
\email{yubosu@princeton.edu}

\begin{abstract}
A widely explored pathway for hot Jupiter (HJ) formation is high-eccentricity migration driven by von Zeipel-Lidov-Kozai cycles induced by an exterior companion.
However, for a distant or low-mass companion, this mechanism typically demands that the planet's initial orbit be very nearly perpendicular to that of the companion. 
In previous work \citep{Yang2025}, we demonstrated that such fine-tuning can be circumvented in the HAT-P-7 system due to the presence of an intermediate body that efficiently couples the orbits of the planet and the distant companion -- a mechanism we termed the \emph{eccentricity cascade} (EC).
In this work, we analytically characterize the dynamics governing the EC and delineate the parameter space within which it effectively operates.
Our qualitative results are as follows:
(i) The proto-HJ's eccentricity is most efficiently excited when the inner triple is on the verge of dynamical instability,
(ii) the addition of a distant fourth body allows this instability to be approached gradually, and
(iii) the instability mechanism is closely related to the stability of circumstellar (S-type) planets in binaries.
By deriving an analytic criterion for S-type instability, we obtain closed-form expressions describing the onset of the EC.
Our results show that efficient HJ formation via the EC occurs across a broad range of intermediate perturbers, highlighting its potential as a robust migration channel.

\end{abstract}

\section{Introduction}

The orbital angular momenta of hot Jupiters (HJs) are frequently misaligned with the spin axes of their host stars \citep{Triaud2010, Albrecht2022}. Such misalignments are commonly interpreted as evidence for high-eccentricity (high-e) migration \citep{Hebrard2008, Winn2010, Dawson2018, Rice2022}, in which a proto-HJ acquires a highly eccentric orbit that subsequently shrinks and circularizes through tidal dissipation. The eccentricity excitation is also often accompanied by inclination excitation, naturally accounting for the observed spin-orbit misalignments.

Among proposed high-eccentricity (high-e) migration pathways, von Zeipel-Lidov-Kozai (ZLK) oscillations are widely regarded as a leading mechanism capable of generating extreme misalignments \citep{Wu2003, Fabrycky2007, Naoz2011, Dawson2014, Petrovich2016, Vick2019, Vick2023, Lu2025}, particularly when driven by distant stellar companions (see \citealp{Albrecht2022} for a recent review). 
However, observational surveys for stellar companions around HJ hosts suggest that only $\sim 10\%$ of HJs could have migrated via ZLK cycles induced by their known stellar companions \citep{Wu2007, Ngo2016}. In most cases, the observed companions are either too distant or too low in mass to drive significant ZLK oscillations \citep{Ngo2016}, or their orbital planes would need to be nearly perpendicular to the planet's initial orbit \citep{Wu2003, Gupta2024}.

\begin{figure*}
    \centering
    \includegraphics[width=0.8\linewidth]{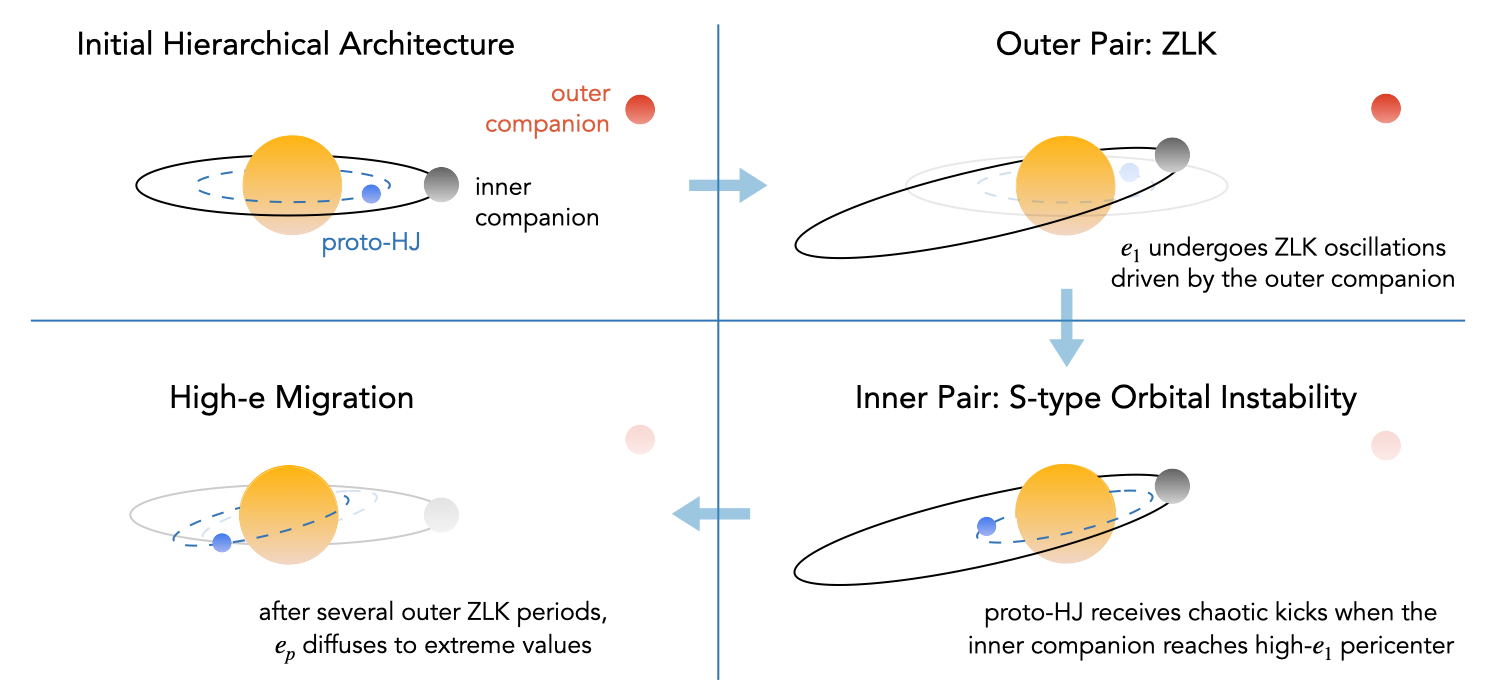}
    \caption{Schematic of HJ formation via the EC mechanism. The planet (blue; subscript ``p'') and inner companion (gray; subscript ``1'') begin on nearly coplanar orbits, while the outer companion (red) is inclined by $\gtrsim 40^\circ$. The outer companion induces ZLK oscillations of the inner companion, periodically exciting its eccentricity and inclination. During high-eccentricity phases, the inner companion repeatedly perturbs the planet through weak scatterings, driving chaotic diffusion in the planet's eccentricity. This process ultimately triggers the planet's migration.}
    \label{fig:ec_cartoon}
\end{figure*}

The HAT-P-7 system exemplifies these challenges: its known stellar companion at a projected separation of 1240~au \citep{Narita2012} is likely too distant to drive effective ZLK migration of the hot Jupiter HAT-P-7b. 
However, in recent work \citep{Yang2025}, we discovered an additional stellar companion at an intermediate separation ($\sim$30~au) through a joint fit to radial-velocity and transit-timing data, and we identified a mechanism involving the additional body that drives high-e migration without stringent inclination constraints.
This mechanism, termed the \emph{eccentricity cascade} (EC), involves a hierarchy of interactions. First, the outer companion periodically drives ZLK oscillations in the inner companion's eccentricity. Then, during its high-e phases, the inner companion repeatedly perturbs the planet through weak scatterings. The cumulative effect of these scatterings, combined with additional secular interactions between the planet and the inner companion, ultimately drive the planet to extreme eccentricities, triggering tidal migration and circularization (see Figures~\ref{fig:ec_cartoon} and \ref{fig:ec_example}). 

The EC has two advantages over standard ZLK formation channels.
First, it facilitates HJ formation even in systems with stellar companions that would otherwise be too distant to drive ZLK migration alone.
Second, it makes no stringent demands on system geometry: the inner companion and planet can initially be nearly coplanar, and the outer companion's inclination need only lie within the ZLK window (i.e., between $40^\circ$ and $140^\circ$ relative to the inner companion).
However, these advantages come at the requirement of an additional body, the inner companion.
The necessary requirements for this body have not been previously characterized, while such an understanding is essential for determining the prevalence of the EC scenario and its applicability towards the broader question of HJ formation.

In this work, we present an analytical model for the EC and derive the necessary properties of the inner companion. 
In Section~\ref{sec:dynamics}, we analyze the scattering dynamics among the inner three bodies that lie at the heart of the EC, deriving new results for the stability of circumstellar (S-type) planets in binary systems.
In Section~\ref{sec:criteria}, we use these results to understand the onset of the EC in the presence of the distant fourth body.
We evaluate the parameter space over which the EC operates and its characteristic timescale, and we validate our results through comparison with \textit{N}-body simulations.
In Section~\ref{sec:discussion}, we relate our findings to previous studies on orbital instability of hierarchical three-body system, assess the viability of the EC in other HJ systems, and comment on the current observational implications of HJ companions.
Hereafter, we denote quantities related to the HJ using the subscript ``p'' (planet), the inner companion with ``1'', and the outer companion with ``2''.

\section{Eccentricity Excitation and the Dynamics of S-type Orbits}\label{sec:dynamics}

The excitation of the planet's eccentricity by the inner companion is the key to the EC.
To study the details of this process, we first neglect the outer companion and consider a simplified configuration where the planet is on an initially circular orbit, perturbed by the inner companion with fixed eccentricity.
Such an architecture has been extensively studied in the context the orbital stability of S-type planets, that is, planets that orbit a host star with a distant binary companion.

Next, we argue that the onset of efficient eccentricity excitation fundamentally depends on the planet's semimajor axis evolution.
In sufficiently hierarchical configurations where the secular approximation applies, eccentricity growth due to a companion remains bounded \citep[e.g.][]{Petrovich2015a}.
Attaining extreme eccentricities therefore requires non-secular planet-companion interactions \citep[e.g.][]{Lee2003}.
These non-secular interactions arise from repeated conjunctions between the two bodies, which ordinarily average to the secular result over long timescales.
However, when each conjunction induces a sufficiently large change in the planet's orbital period (equivalently, its semimajor axis), the sequence of conjunctions becomes phase-incoherent rather than periodic, breaking the secular approximation \citep[an analogous threshold arises in the tidal circularization of highly-eccentric HJs; see][]{Wu2018, Vick2019}.
Thus, determining when the planet's semimajor axis evolves significantly enough to decorrelate successive conjunctions provides the key criterion for the onset of efficient eccentricity excitation.

In this section, we show that the semimajor axis evolution of the planet can be understood analytically via the so-called Chirikov standard map \citep{Chirikov1979}. We use this correspondence to identify conditions under which orbital diffusion sets in, marking the onset of significant eccentricity growth. Our analysis yields a closed-form analytical stability criterion for S-type planetary orbits.

\subsection{The map}\label{ssec:the_map}
We consider the dynamics of a planet on an initially circular orbit, perturbed by an exterior stellar companion (the inner companion in the EC) with fixed eccentricity.
Each time the companion passes through its pericenter, it induces a ``kick'' to the planet's semimajor axis. In general, this kick function can be a complex and sharply-peaked function of the planet's orbital phase \citep{Hadden2024}.
However, for sufficiently distant pericenter passages, the net change to the planet's semimajor axis is approximately a sinusoidal function of the planet's instantaneous orbital phase when the companion passes pericenter:
\begin{equation}
    a_p' = a_p + \delta a_{p,\max}\sin{f},\label{eq:iterative_map_a}
\end{equation}
where $a_p$ and $a_p'$ are the planet's semimajor axes at successive pericenter passages of the companion, $\delta a_{p,\max}$ is the maximum semimajor-axis change over one companion orbit, and $f$ is the angular separation between the planet and the companion when the latter is at pericenter.

Between successive passages, the angular separation evolves by $\Delta f=2\pi n_p'/n_1$, where $n_p'=\sqrt{GM_\star/a_p'^3}$ and $n_1=\sqrt{G(M_\star+m_1)/a_1^3}$ are the mean motions of the planet and the companion, respectively.
This can be rewritten as
\begin{equation}
    f' = f + 2\pi \sqrt{1-\mu} \left(\frac{a_1}{a_p'}\right)^{3/2} \quad {\rm (mod\, 2\pi)},
    \label{eq:iterative_map_f}
\end{equation}
where $\mu \equiv m_1/(M_\star+m_1)$ is the binary mass ratio.

By expanding Equation~\eqref{eq:iterative_map_f} about the planet's initial semimajor axis $a_{p,0}$ and non-dimensionalizing, we can combine Equations~\eqref{eq:iterative_map_a} and \eqref{eq:iterative_map_f} into an iterative map:
\begin{equation}
\begin{aligned}
    A' &= A - K\sin f, \\
    f' &= f  + A' + 5\pi \sqrt{1-\mu} (a_1/a_{p,0})^{3/2} \quad {\rm (mod\, 2\pi)},
    \label{eq:standard_map}
\end{aligned}
\end{equation}
where we define $A \equiv - 3\pi \sqrt{1-\mu}\,{a_1^{3/2}}{a_{p,0}^{-5/2}}a_p$ and analogously for $A'$, and the stochasticity parameter $K$ is
\begin{equation}
    K = 3\pi \sqrt{1-\mu} \left(\frac{a_1}{a_{p,0}}\right)^{3/2} \, \frac{\delta a_{p,\max}}{a_{p,0}}.
    \label{eq:K_prelim}
\end{equation}

For a planet initially on a circular, coplanar orbit with the companion\footnote{We justify the assumption of coplanarity as a general condition for the EC in Section~\ref{ssec:geometry}.}, detailed analytical calculations (see Appendix~\ref{app:delta_a}) yield
\begin{equation}
    \frac{\delta a_{p,\max}}{a_{p,0}} \approx \zeta \Exp{-{\tfrac{2n_{p,0}}{n_1}\left(\cosh^{-1}\left(\tfrac{1}{e_1}\right) -j_1\right)}},
\end{equation}
where $e_1$ is the eccentricity of the companion, $j_1 \equiv \sqrt{1-e_1^2}$, and $\zeta$ is a dimensionless parameter given by
\begin{equation}
\begin{aligned}
    \zeta = &~ \frac{90\pi^2}{e_1^4} \left(\frac{m_1}{m_{\rm tot}}\right)^2 
     \left(1+\frac{n_1}{n_{p,0}}\frac{e_1^2}{3j_1^3}\right)
    \left(1+\frac{n_1}{n_{p,0}}\frac{6+9e_1^2}{5j_1^3} \right),
\end{aligned}
\end{equation}
where $\frac{m_1}{m_{\rm tot}} = \frac{m_1}{M_\star+m_p+m_1} \approx \mu$.

Combining the expressions above, we get
\begin{equation}
\begin{aligned}
    K = &~ 3\pi\sqrt{1-\mu} \left(\frac{a_1}{a_{p,0}}\right)^{3/2} \zeta \\
    &~ \times \Exp{-{\tfrac{2n_{p,0}}{n_1}\left(\cosh^{-1}\left(\tfrac{1}{e_1}\right) -j_1\right)}}.
    \label{eq:K}
\end{aligned}
\end{equation}
This expression, in conjunction with Equations~\eqref{eq:standard_map}, provides a model for the long-term evolution of the planet's semimajor axis in the form of a simple iterative map.

\subsection{Onset of chaotic diffusion}\label{ssec:instab}
The onset of chaotic diffusion in a system described by Equation~\eqref{eq:standard_map} is governed by the stochasticity parameter $K$. 
In the classical Chirikov standard map, global chaos emerges once $K$ exceeds a critical threshold, $K_{\rm crit}$, of order unity \citep[e.g.][]{ott2002_chaosdynamicalsystemsbook}.
However, given that the map serves only as an approximate model for the \textit{N}-body dynamics considered here, we investigate the correlation between the value of $K$ and the emergence of chaotic diffusion numerically via \textit{N}-body simulations.

We initialize a Jupiter-mass planet on a circular orbit, coplanar with an exterior companion. The companion's argument of pericenter is held fixed, while we sample 1000 parameter sets from
$a_p\!\in[3,7]\,\mathrm{au}$, $P_1\!\in[100,300]\,\mathrm{yr}$, $m_1\!\in[0.05,0.5]\,M_\odot$, and $e_1\!\in[0.5,0.9]$.
For each configuration, we further randomly sample 10 different initial anomalies of both bodies.
Then, we evolve each system for $10^6 P_1$ using the \texttt{IAS15} integrator \citep{ias15} in \texttt{REBOUND} \citep{rebound}.

For systems in which chaotic diffusion occurs, we define the instability timescale as the median time -- across the ensemble of initial anomalies -- for the planet's eccentricity to diffuse to $e_p = 0.99$. In the context of high-e HJ migration, this threshold corresponds to the onset of efficient tidal circularization. 

\begin{figure}
    \centering
    \includegraphics[width=0.85\linewidth]{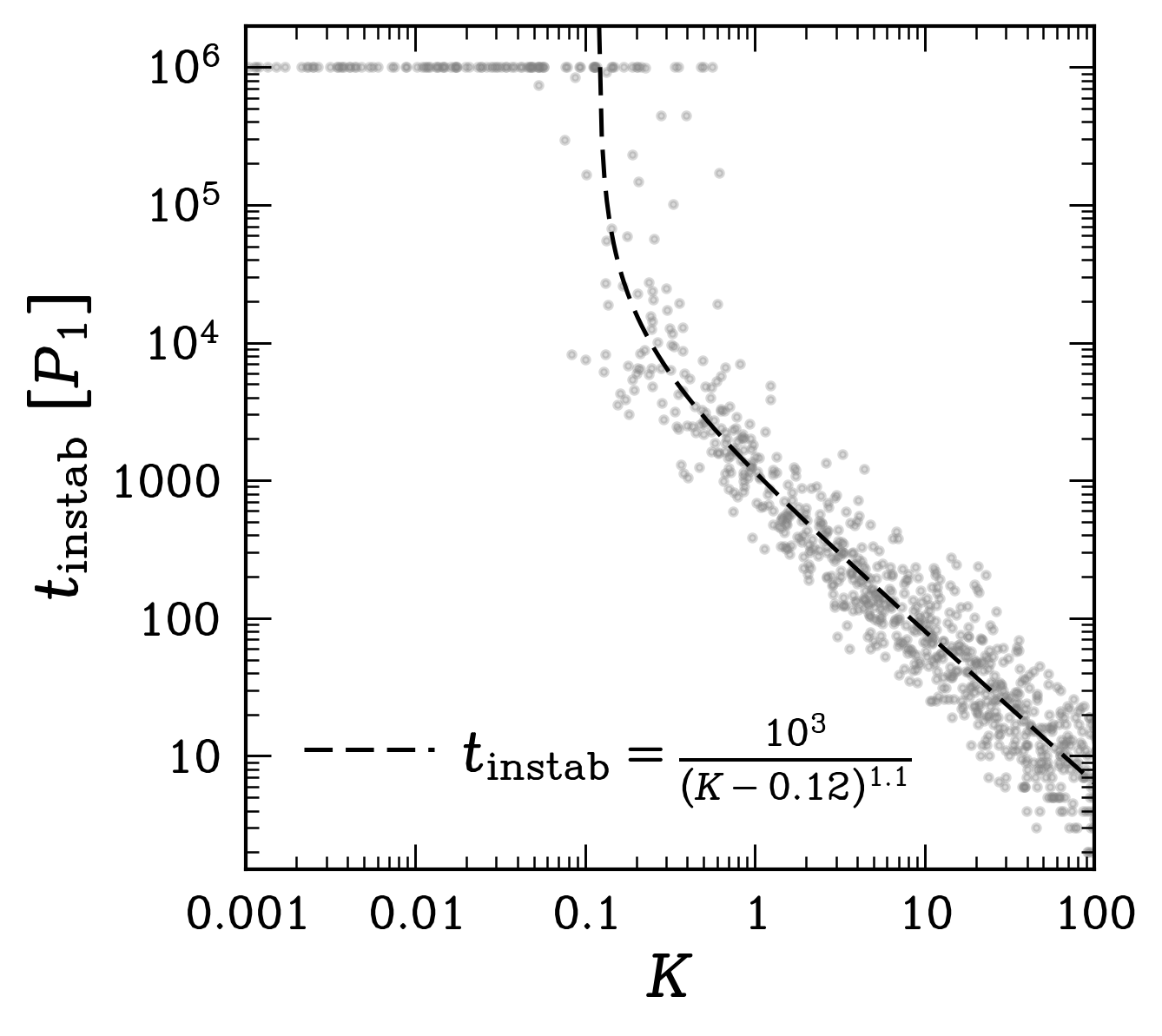}
    \caption{Instability timescale for the planet's eccentricity to reach an extreme value of $e_p=0.99$ as a function of the stochasticity parameter $K$ defined in Equation~\eqref{eq:K}. Systems that remain stable for the full integration ($10^6 P_1$) appear as a ceiling in the distribution.}
    \label{fig:T_vs_K}
\end{figure}

Figure~\ref{fig:T_vs_K} shows the instability timescale as a function of $K$. Systems that remain stable throughout the integration appear as a ceiling at $10^6P_1$. A fit to the data indicates that instability only arises once $K$ exceeds a critical value $K_{\rm crit} \approx 0.12$\footnote{We remark that this value is approximately an order of magnitude lower than that from the classical Chirikov standard map.
This may be due to our approximation of the kick function in Equation~\eqref{eq:iterative_map_a} as a simple sinusoid: in general, the kick function is periodic and can be expressed as a sine series in $f$. Our approximation neglects the higher harmonics (i.e., $\sin{nf}$ for $n>1$), while such higher-order terms may result in resonance overlap that enhances chaos and reduces $K_{\rm crit}$ \citep{Hadden2024}.}.
Beyond this threshold, the instability timescale scales tightly with $K$. Thus, the stochasticity parameter $K$ provides a predictive criterion for the long-term stability boundary and governs the dynamical evolution of S-type orbits.

We next translate this critical stability threshold into a condition on the orbital architecture. Specifically, we define the critical eccentricity of the companion, $e_{1,\rm crit}$, as the eccentricity for which the stochasticity parameter equals the diffusion threshold, i.e., $K(e_{1,\rm crit})=0.12$. 
Near this threshold, where $n_1/n_p \ll j_1^3/e_1^2$, Equation~\eqref{eq:K} yields the approximate condition
\begin{equation}
\begin{aligned}
    &~ e_{1,\rm crit}^{-4}\Exp{-{\tfrac{2n_p}{n_1}\left(\cosh^{-1}\left(\tfrac{1}{e_{1,\rm crit}}\right) - j_{1,\rm crit}\right)}} \\
    &~ \approx \frac{K_{\rm crit}}{270\pi^3}\frac{1}{\mu^2\sqrt{1-\mu}} \left(\frac{a_p}{a_1}\right)^{3/2}.
    \label{eq:e_crit}
\end{aligned}
\end{equation}
If the companion's eccentricity remains below $e_{1,\rm crit}$, the planet avoids entering the chaotic regime and remains dynamically stable. 
Note that, as $e_1\to 1$, the exponent in Equation~\eqref{eq:e_crit} asymptotically approaches $\exp\left(-\frac{4\sqrt2}{3}\frac{n_p}{\omega_{1,\rm peri}}\right)$, where $\omega_{1,\rm peri} \equiv \sqrt{G (M_\star + m_1) / [a_{1}(1 - e_1)]^3}$ is the angular frequency of the companion at its pericenter.

\begin{figure}
    \centering
    \includegraphics[width=1\linewidth]{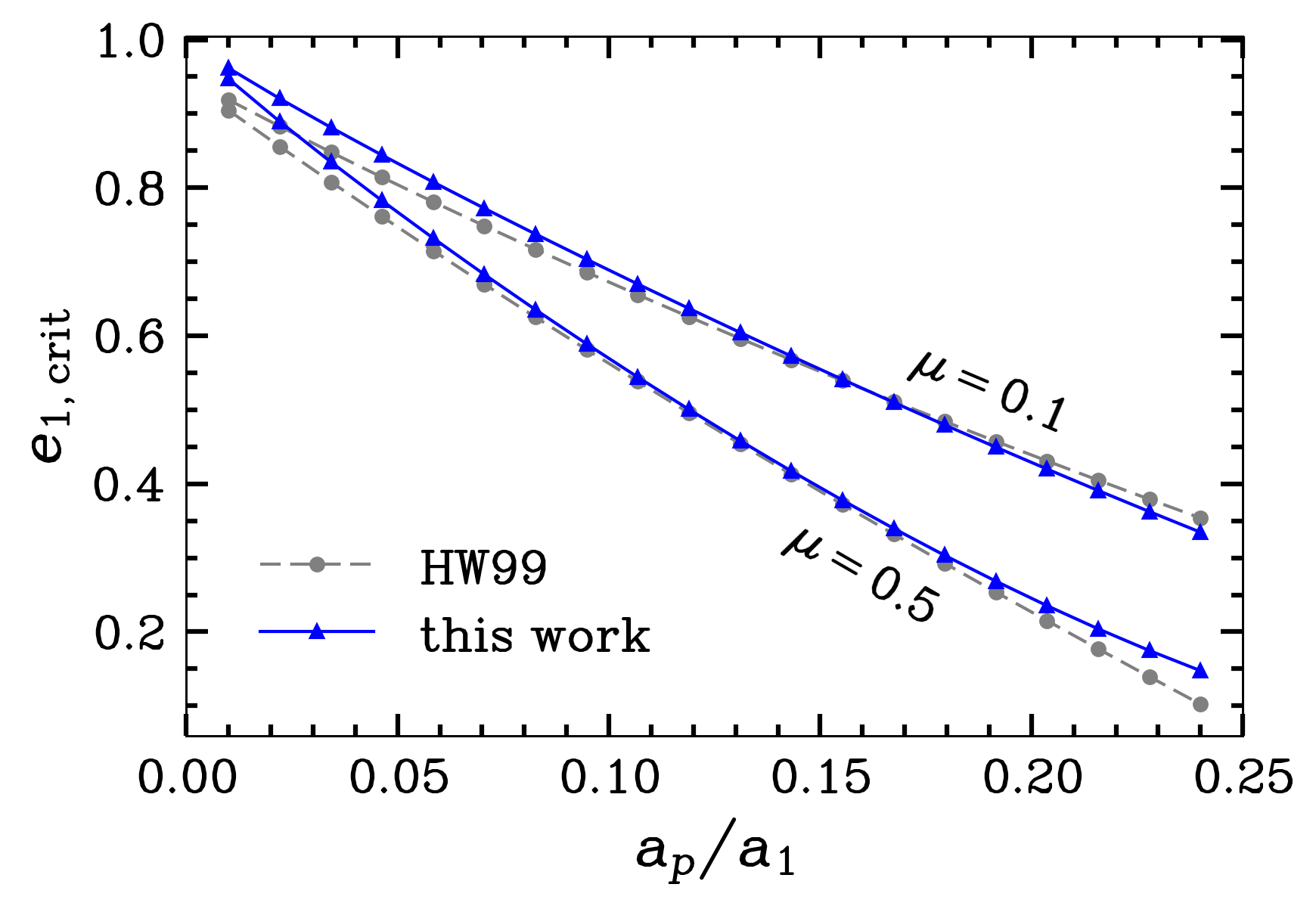}
    \caption{Comparison between our analytical stability criterion (Equation~\ref{eq:e_crit}) and the empirical stability boundary from HW99. The figure shows the critical companion eccentricity $e_{1,\mathrm{crit}}$ as a function of the semimajor axis ratio $a_p/a_1$ for binary mass ratios $\mu = 0.1$ and $\mu = 0.5$.}
    \label{fig:stab_boundary}
\end{figure}

In Figure~\ref{fig:stab_boundary}, we compare our analytical stability criterion (Equation~\ref{eq:e_crit}) with the empirical stability boundary from \citet[hereafter HW99]{Holman1999}.
Results are shown for binary mass ratios $\mu=0.1$ and $0.5$, chosen to span the typical regime from low-mass to equal-mass stellar companions.
The analytical and empirical boundaries agree well across a wide range of binary separations.
This result complements the established understanding of S-type orbital stability and may be extended to characterize the stability of more general hierarchical triple systems (see Section~\ref{ssec:disc_stability}).

Finally, we remark that the instability timescales in Figure~\ref{fig:T_vs_K} are well described by the empirical fit shown as the black dashed line, given by
\begin{equation}
    t_{\rm instab} \approx \frac{10^3P_1}{(K-K_{\rm crit})^{1.1}}.
    \label{eq:t_D}
\end{equation}

\section{Onset of the eccentricity cascade}\label{sec:criteria}
The previous section focused on the dynamics between the planet and the inner companion in isolation. We now reintroduce the outer companion in the EC scenario and apply those results to understand the planet's eccentricity evolution under the presence of both companions.

\subsection{Geometric assumptions}\label{ssec:geometry}
The analytic framework developed in the previous section assumes that the orbits of the planet and the inner companion are initially coplanar. In the absence of an outer perturber, the coplanarity is preserved. Below, we justify this assumption and demonstrate that it still holds even under the perturbations by the outer companion.

First, initial coplanarity between the planet and the inner companion is naturally expected. Planet formation theory generally predicts minimal orbital misalignment between a planet and a closely orbiting stellar companion \citep{Lubow_2000, Foucart_2013}. Observationally, planetary orbits are found to be preferentially aligned with close-in companion stars \citep{Christian2022, Dupuy2022, Zhang2025}\footnote{Although \citep{Christian2025} found the evidence for coplanarity to be weaker or absent for gas giant planets, their sample of giants was dominated by HJs, which likely experienced significant orbital evolution from their initial configurations.}.

Second, orbital coplanarity between the planet and the inner companion is maintained throughout the EC due to strong dynamical coupling. Specifically, the orbital precession of the inner companion induced by the outer companion is slow compared to that of the planet induced by the inner companion. This condition can be quantified as:
\begin{equation}
\frac{\Omega_{12}}{\omega_{p1}} \simeq \frac{m_2}{m_1}\frac{a_1^{9/2}}{a_2^3,a_p^{3/2}} \ll 1,
\label{eq:Omega}
\end{equation}
where $\Omega_{12}$ denotes the orbital precession frequency between the inner and outer companions, and $\omega_{p1}$ represents the precession frequency between the planet and inner companion \citep{Lai2017}. 
Although violation of the condition in Equation~\eqref{eq:Omega} does not preclude HJ formation (see Section~\ref{ssec:necessary_criteria}), satisfaction of the condition ensures that the planet and the inner companion remain nearly coplanar throughout the EC, even as the orbital orientation of the inner companion changes due to ZLK oscillations driven by the outer companion (see, e.g., the bottom panel of Figure~\ref{fig:ec_example}).

\subsection{Maximum \texorpdfstring{$e_1$}{e1} in ZLK cycles}\label{ssec:emax}

In addition to affecting the inner companion's inclination, the outer companion also modulates its eccentricity through the ZLK mechanism. 
A key question for the EC theory is whether these ZLK oscillations can drive the inner companion's eccentricity beyond the critical threshold $e_{1,\rm crit}$ (derived in Section~\ref{ssec:instab}), which is required to trigger chaotic diffusion in the planet's orbit. In this section, we first assess the range of eccentricities attainable by ZLK oscillations driven by the outer companion.

The maximum eccentricity attained by the inner companion depends on the strength of other sources of apsidal precession, which act to suppress the growth of eccentricity \citep{Liu2015}. In the present configuration, the dominant source arises from the quadrupole potential of the planet's orbit. 
Neglecting octupole-order interactions between the companions, the maximum eccentricity, $e_{1,\max}$, is determined by conservation of the inner companion's angular momentum along the outer companion's orbit normal, $j_1\cos{i_{12}}$, together with conservation of its total potential
\begin{equation}
    \langle{\Phi_{\rm tot}}\rangle \approx \langle{\Phi_{\rm quad}}\rangle + \langle{\Phi_{\rm orb}}\rangle,
\end{equation}
where the interaction between the companions is described by the double averaged quadrupole potential
\begin{equation}
\begin{gathered}
    \langle{\Phi_{\rm quad}}\rangle = -\frac{GM_\star m_1 m_2}{8(M_\star+m_1)}\frac{a_1^2}{a_2^3(1-e_2^2)^{3/2}}\\
    \times\left[2+3e_1^2-(3+12e_1^2-15e_1^2\cos^2\omega_1)\sin^2{i_{12}}\right],
\end{gathered}
\end{equation}
and the apsidal precession driven by the planet is described by the potential
\begin{equation}
    \langle{\Phi_{\rm orb}}\rangle =  - \frac{G m_1m_pa_p^2}{4 a_1^3 (1-e_1^2)^{3/2}}.
    \label{eq:phi_orb}
\end{equation}
Conservation of $j_1 \cos i_{12}$ and $\langle \Phi_{\rm tot} \rangle$ leads to an implicit equation for the maximum eccentricity of the inner companion attained in ZLK cycles:
\begin{equation}
\begin{aligned}
    \frac{\epsilon}{3} \left(\frac{1}{j_{1,\min}^{3}} - \frac{1}{j_{1,0}^{3}}\right) \\
    =  \frac{9}{8}(e_{1,\max}^2-e_{1,0}^2) + \frac{15}{8} e_{1,0}^2\cos^2{\omega_{1,0}}\sin^2{i_{12,0}} \\
    - \frac{3}{8}\left[\frac{(1+4e_{1,\max}^2)j_{1,0}^2}{j_{1,\min}^2} - (1+4e_{1,0}^2)\right] \cos^2{i_{12,0}},
    \label{eq:e1_max}
\end{aligned}
\end{equation}
where $j_{1,\min}\equiv \sqrt{1-e_{1,\max}^2}$, and analogously for $j_{1,0}$. The dimensionless parameter, $\epsilon$, is defined as
\begin{align}
    \epsilon 
        \equiv{}&
            \frac{3}{4}
            \frac{(M_\star+m_1)m_p}{M_\star m_2}
            \frac{a_p^2a_2^3}{a_1^5}
            (1-e_2^2)^{3/2}\nonumber\\
        \approx{}&
            0.65
            \left(1 + \frac{m_1}{M_\star}\right)
            \left(\frac{m_p}{1.84 M_{\rm J}}\right)
            \left(\frac{m_2}{0.15 M_{\odot}}\right)^{-1}
            \left(\frac{a_p}{3\;\mathrm{au}}\right)^2\nonumber\\
        &\times
            \left(\frac{a_2\sqrt{1 - e_2^2}}{521\;\mathrm{au}}\right)^3
            \left(\frac{a_1}{28\;\mathrm{au}}\right)^{-5},
\end{align}
where the fiducial parameters in the last two lines are the observed parameters of the HAT-P-7 system \citep{Yang2025}.
Qualitatively, $\epsilon \lesssim 1$ yields large eccentricity excitation of the inner companion.

We note that the results derived above hold exactly for ZLK cycles at quadrupole order.
When octupole-order effects are included, however, the orbital parameters $e_{1}$, $\omega_{1}$, and $i_{12}$ at the start of each ZLK cycle generally do not return exactly to their initial values, but instead vary chaotically on a timescale much longer than the ZLK oscillation period. 
This longer modulation timescale, often referred to as the octupole timescale or the timescale of the eccentric Kozai-Lidov mechanism \citep[EKM, e.g.,][]{Naoz2016}, is given by \citet{Antognini2015} as
\begin{equation}
    t_{\rm EKM} \approx \frac{256\sqrt{10}}{15\pi\sqrt{\epsilon_{\rm oct}}}t_{\rm ZLK},
    \label{eq:t_EKM}
\end{equation}
where $\epsilon_{\rm oct} = \frac{M_\star-m_1}{M_\star+m_1} \frac{a_1}{a_2} \frac{e_2}{1-e_2^2}$, and the ZLK timescale is
\begin{gather}
    t_{\rm ZLK} \approx \frac{16}{15}\left(\frac{a_2^3}{a_1^{3/2}}\right)\sqrt{\frac{M_\star+m_1}{Gm_2^2}} (1-e_2^2)^{3/2}. 
    \label{eq:t_ZLK}
\end{gather}

In other words, the quantities $e_{1,0}$, $\omega_{1,0}$ and $i_{12,0}$ are no longer constant over the octupole timescale $t_{\rm EKM}$. As a result, Equation~\eqref{eq:e1_max} must be modified to account for these octupole-order variations.
As noted by \citet{huang2025_kozai}, the inner companion's maximum eccentricity is capped by a limiting value $e_{1,\lim}$, obtained by setting $i_{12,0}=90^\circ$ in Equation~\eqref{eq:e1_max}.
Specifically,
\begin{equation}
    \frac{\epsilon}{3} \left(\frac{1}{j_{1,\lim}^{3}} - \frac{1}{j_{1,0}^{3}}\right) \\
    =  \frac{9}{8}e_{1,\lim}^2 + \frac{3}{4}e_{1,0}^2,
    \label{eq:e1_lim}
\end{equation}
where $j_{1,\lim} \equiv \sqrt{1-e_{1,\lim}}$.
In general, $e_{1,\lim}$ provides an estimate of the maximum eccentricity attainable by the inner companion for a given $e_{1,0}$ once octupole-order ZLK effects are included, assuming the system is allowed to evolve for multiple $t_{\rm EKM}$ cycles.

\subsection{Necessary criteria}\label{ssec:necessary_criteria}
Now that we have characterized the maximum eccentricity attainable by the inner companion during ZLK cycles, we use it to delineate the parameter regime over which the EC operates.
Conceptually, the EC sets in when the inner companion's eccentricity grows sufficiently large to trigger chaotic diffusion in the planet's orbit.
Since the maximum attainable eccentricity $e_{1,\lim}$ for a given architecture is given by Equation~\eqref{eq:e1_lim}, the onset of the EC requires
\begin{equation}
    e_{1,\lim}^{\rm (ZLK)} \geq e_{1,\rm crit}^{\rm (diff)},
    \label{eq:criteria}
\end{equation}
where $e_{1,\rm crit}^{\rm (diff)}$ is given by Equation~\eqref{eq:e_crit}.

Directly validating Equation~\eqref{eq:criteria} through numerical integrations, however, is challenging because the time required for the inner companion to reach $e_{1,\lim}$ via chaotic octupole-level evolution can be arbitrarily long.
In order to study the long-timescale behavior of the system without unnecessarily expensive integrations, we begin by selecting favorable initial conditions such that $e_{1,\lim}$ is reached during the first ZLK cycle.
Practically, this limiting configuration is realized by initializing the system with $i_{12,0}=90^\circ$ in \textit{N}-body simulations.
We reiterate that, although many systems with more general values of $i_{12,0}$ still readily reach $e_{1,\lim}$ through octupole-order evolution, our chosen setup ensures a transparent and computationally efficient comparison between the analytical criterion and the numerical results.
We will later extend our analysis to include more representative initial conditions.

\begin{figure}
    \centering
    \includegraphics[width=1\linewidth]{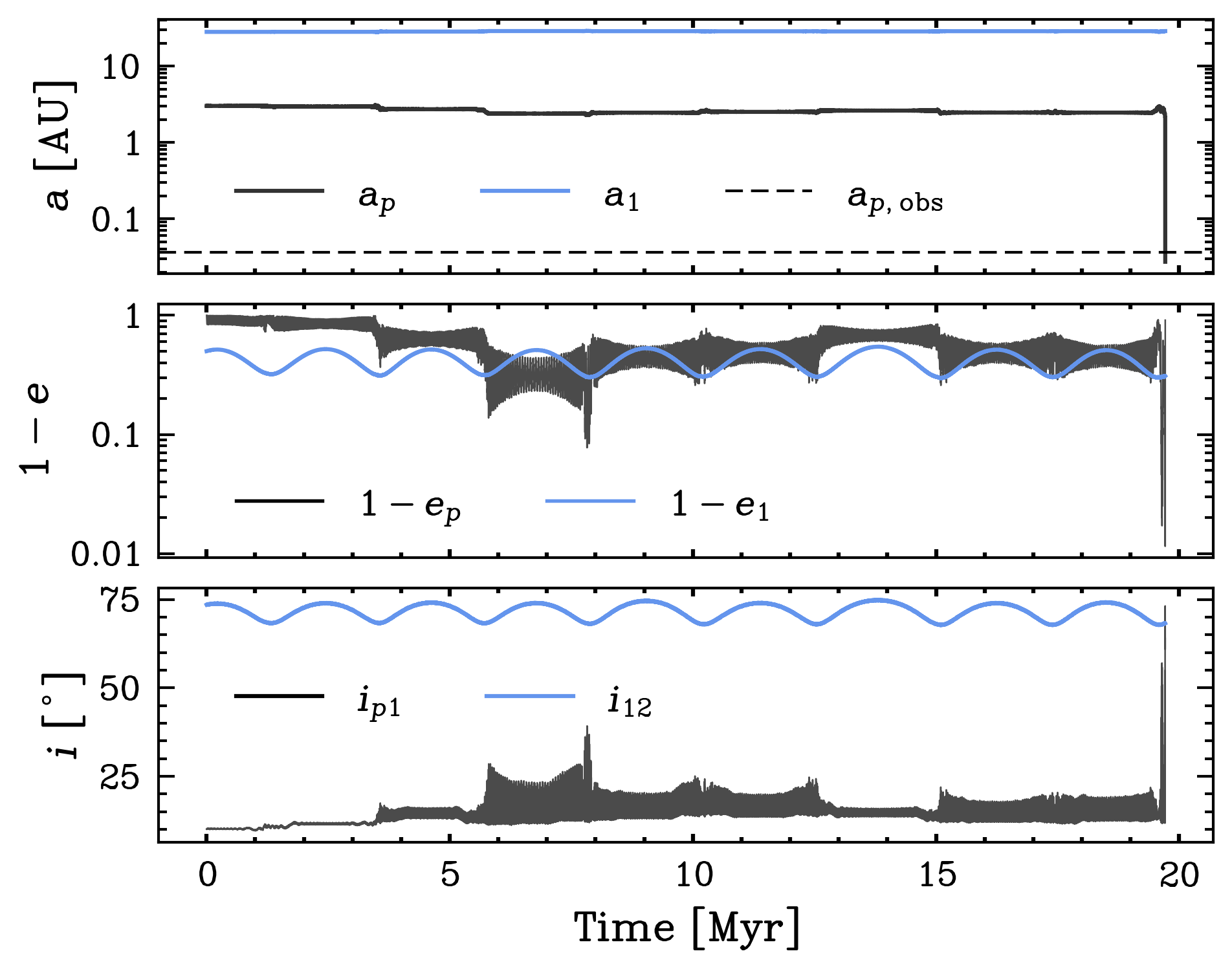}
    \caption{Example of HJ formation in the HAT-P-7 system via the EC mechanism, adapted from \cite{Yang2025}. The three panels show the evolution of semimajor axes, eccentricities and mutual inclinations. The dashed line in the top panel denotes the present-day orbital location of HAT-P-7b.}
    \label{fig:ec_example}
\end{figure}

\begin{figure}
    \centering
    \includegraphics[width=1\linewidth]{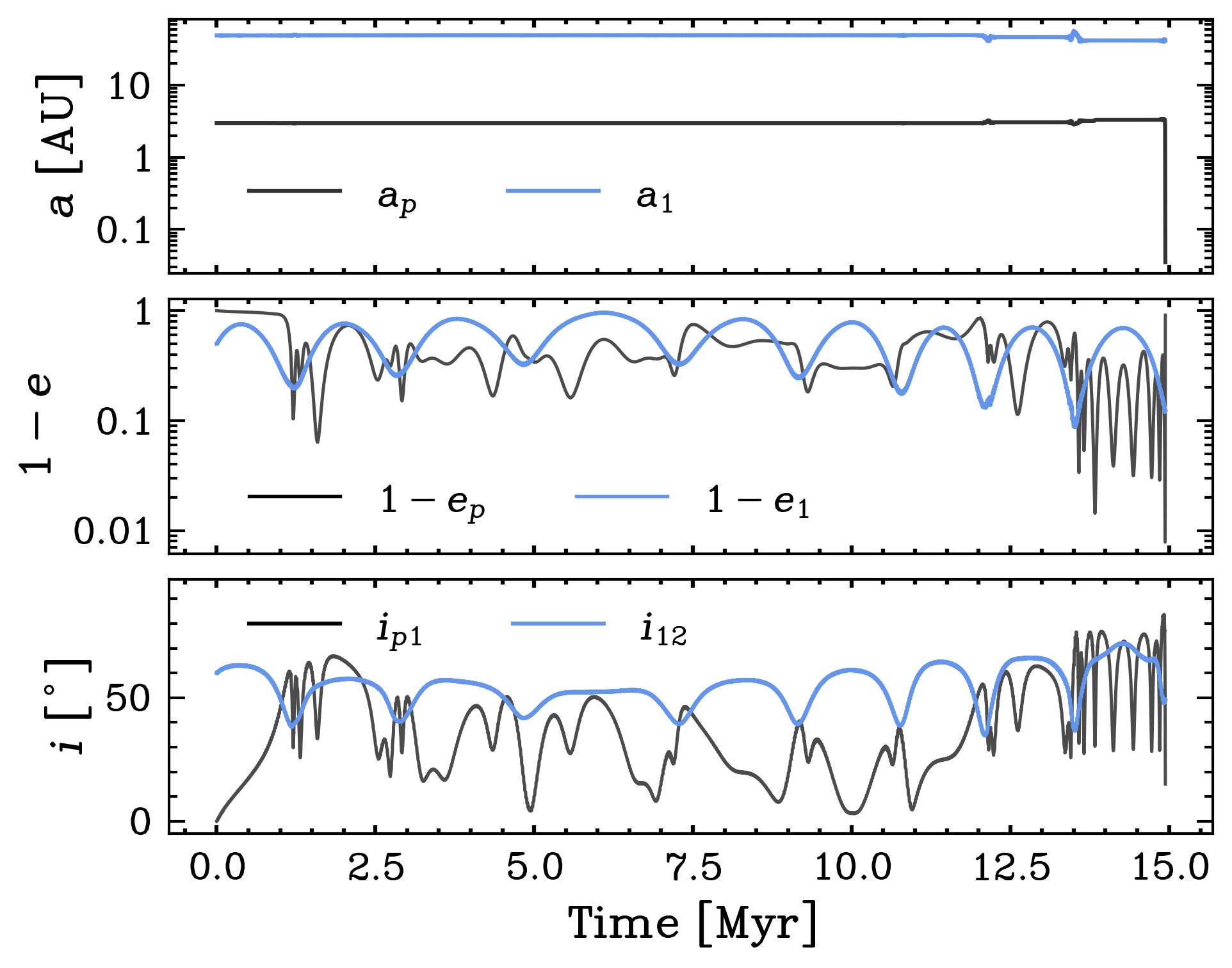}
    \caption{Same system configuration as in Figure~\ref{fig:ec_example}, but with a lower-mass inner companion on a wider orbit, resulting in $\Omega_{12}/\omega_{p1}=0.3$ (compared to 0.001 previously). The planet and inner companion are no longer strongly coupled, allowing their mutual inclination to increase as the inner companion undergoes ZLK oscillations induced by the outer companion. Around 13~Myr, the mutual inclination exceeds the ZLK threshold, initiating high-e migration of the planet via the classical ZLK mechanism.}
    \label{fig:Kozai-Kozai}
\end{figure}

In this restricted setup, we identify four distinct dynamical regimes governed by the properties of the inner companion:
\begin{enumerate}[{(}1{).}]
    \item If $\Omega_{12}/\omega_{p1} \gtrsim 0.1$, the planet and the inner companion are no longer dynamically coupled, and the conditions for the EC are no longer met.
    However, formation via other mechanisms can take place (see Figure~\ref{fig:Kozai-Kozai}).
    \item If $e_{1,\lim}^{\rm (ZLK)} \geq e_{1,\rm crit}^{(\rm diff)}$, the system can form an HJ via the EC, with an efficiency depending on the inner companion's initial orientation (i.e., inclination and argument of pericenter) relative to the outer companion. An example evolution is shown in Figure~\ref{fig:ec_example}.
    \item If $e_{1,\lim}^{\rm (ZLK)} < e_{1,\rm crit}^{(\rm diff)}$, the inner companion cannot excite the planet beyond the diffusion threshold, and high-e migration is not expected.
    \item If $e_{1,0} \geq e_{1,\rm crit}^{(\rm diff)}$, the system is initially unstable, and the planet may undergo rapid migration, collision with the host star, or ejection.
\end{enumerate}

Next, we apply these results to a hypothetical system modeled after HAT-P-7. Specifically, we adopt the same fiducial parameters for the host star, planet, and outer companion as in \citet[Section~3.1]{Yang2025}. For the inner companion, rather than fixing its properties to the observed constraints, we explore a range of its mass and semimajor axis, while we assume coplanarity with the planet's orbit and fix its initial eccentricity to be $e_{1,0}=0.5$.

\begin{figure*}
    \centering
    \includegraphics[width=0.9\linewidth]{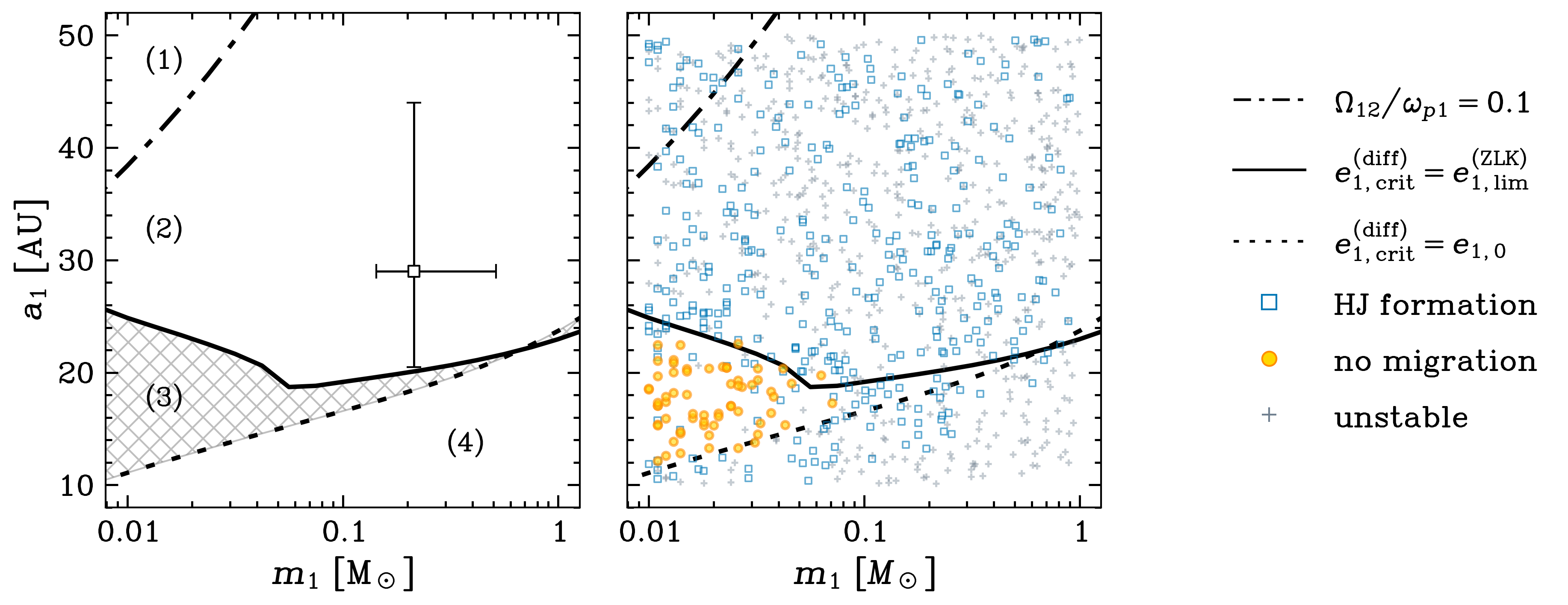}
    \caption{Dynamical regimes for the EC in a HAT-P-7-like system, shown as a function of the inner companion's mass and semimajor axis. We assume that the inner companion has an initial eccentricity of $e_{1,0}=0.5$ and eventually reaches $e_{1,\lim}$. \textbf{Left:} The parameter space is divided into four regions by the analytical boundaries: (1) where the planet and inner companion become dynamically decoupled, and alternative migration pathways may operate; (2) where HJ formation via the EC is viable, with efficiency dependent on the inner companion's initial orientation; (3) where the EC is strongly suppressed; and (4) where the planet is expected to undergo rapid migration, collision, or ejection due to immediate instability. The data point marks the observation constraints on the inner companion's orbit, assuming that the planet formed at 3~au and applying the long-term stability requirements \citep{Yang2025}.
    \textbf{Right:} Outcomes of 1000 direct \textit{N}-body simulations over 50~Myr using the parameters listed in Table~\ref{tab:param}, with $i_{12,0}=90^\circ$. The data points represent systems where the planet becomes an HJ (blue), does not migrate (orange), or undergoes tidal disruption, collision, or ejection (gray). The tidal disruption radius is set to $2R_p(M_\star/m_p)^{1/3}=0.013$~au \citep{Guillochon2011}.}
    \label{fig:HAT-P-7_param_space}
\end{figure*}

Figure~\ref{fig:HAT-P-7_param_space} compares the analytical predictions with simulations. The left panel shows the analytically derived boundaries for the onset of the EC (see legend) assuming $e_{1,\max}=e_{1,\lim}$. The four dynamical regimes identified above are labeled, and the data point marks the location of the observationally identified inner companion in the HAT-P-7 system \citep{Yang2025}.
The right panel overlays these analytical boundaries with the outcomes of 1000 direct \textit{N}-body simulations. The full simulation setup is summarized in Table~\ref{tab:param}, and each system is evolved for $50$~Myr\footnote{The ZLK oscillation period ranges from $\sim$0.5 to $\sim$5~Myr across the sampled parameter space, making 50~Myr a sufficient integration timescale.}, with outcomes indicated by the symbols in the legend.

\begin{table}[t]
    \centering
    \caption{Adopted parameters for the \texttt{REBOUND} simulations of a HAT-P-7–like system. From top to bottom, the quantities listed are: masses, eccentricities, semimajor axes, true anomalies, radii, tidal Love numbers, tidal quality factors, and spin frequencies.}
    \begin{threeparttable}
        \begin{tabularx}{\columnwidth}{l l}
            \hline
            Parameters & Values\\
            \hline
            $M_\star$, $m_p$ $m_1$, $m_2$ & 
            $1.35\,\rm M_\odot$, $1.84\,\rm M_J$, $\in[0.01,1]\,\rm M_\odot$, $0.15\,\rm M_\odot$\\
            $e_p$, $e_1$, $e_2$ & 0, 0.5, 0.7\\
            $a_p$, $a_1$, $a_2$ & 3\,au,  $\in [10,50]$\,au, 730\,au\\
            $f_p$, $f_1$, $f_2$ & $\in[0,2\pi]$ \\
            $R_\star$, $R_p$ & $1.99\,\rm R_\odot$, $16.9\,\rm R_\oplus$\\
            $k_{2\star}$, $k_{2p}$ & 0.03, 0.5\\
            $Q_{\star}, Q_p$ & $10^7$, $10^4$\\
            $\Omega_\star$, $\Omega_p$ & $2\pi/(7\,\rm d)$, $2\pi/(10\,\rm hr)$\\
            \hline
        \end{tabularx}
    \end{threeparttable}
    \label{tab:param}
\end{table}

In Region (2) where $a_1$ is large, the limiting eccentricity $e_{1,\lim}$ is large due to the stronger perturbations from the outer companion and the weaker apsidal precession induced by the planet. In this regime, $e_{1, \rm crit}$ is always exceeded, and the inner system undergoes dynamical instability, triggering HJ formation via the EC. As $a_1$ decreases into Region (3), where $e_{1,\lim}<e_{1,\rm crit}$, the EC is strongly suppressed, and migration is mostly absent (orange points in the right panel). For even smaller $a_1$ in Region (4), $e_{1,\lim}$ falls below the initial eccentricity $e_{1,0}$, making the system immediately unstable even without ZLK oscillations. Planets in such systems typically undergo rapid ejection rather than migration, which requires at least one secular cycle. Finally, while systems in Region (1) fall outside the EC regime due to dynamically decoupling of the inner pair, HJ formation still occurs frequently through alternative pathways (see Figure~\ref{fig:Kozai-Kozai}).

\begin{figure}
    \centering
    \includegraphics[width=0.75\linewidth]{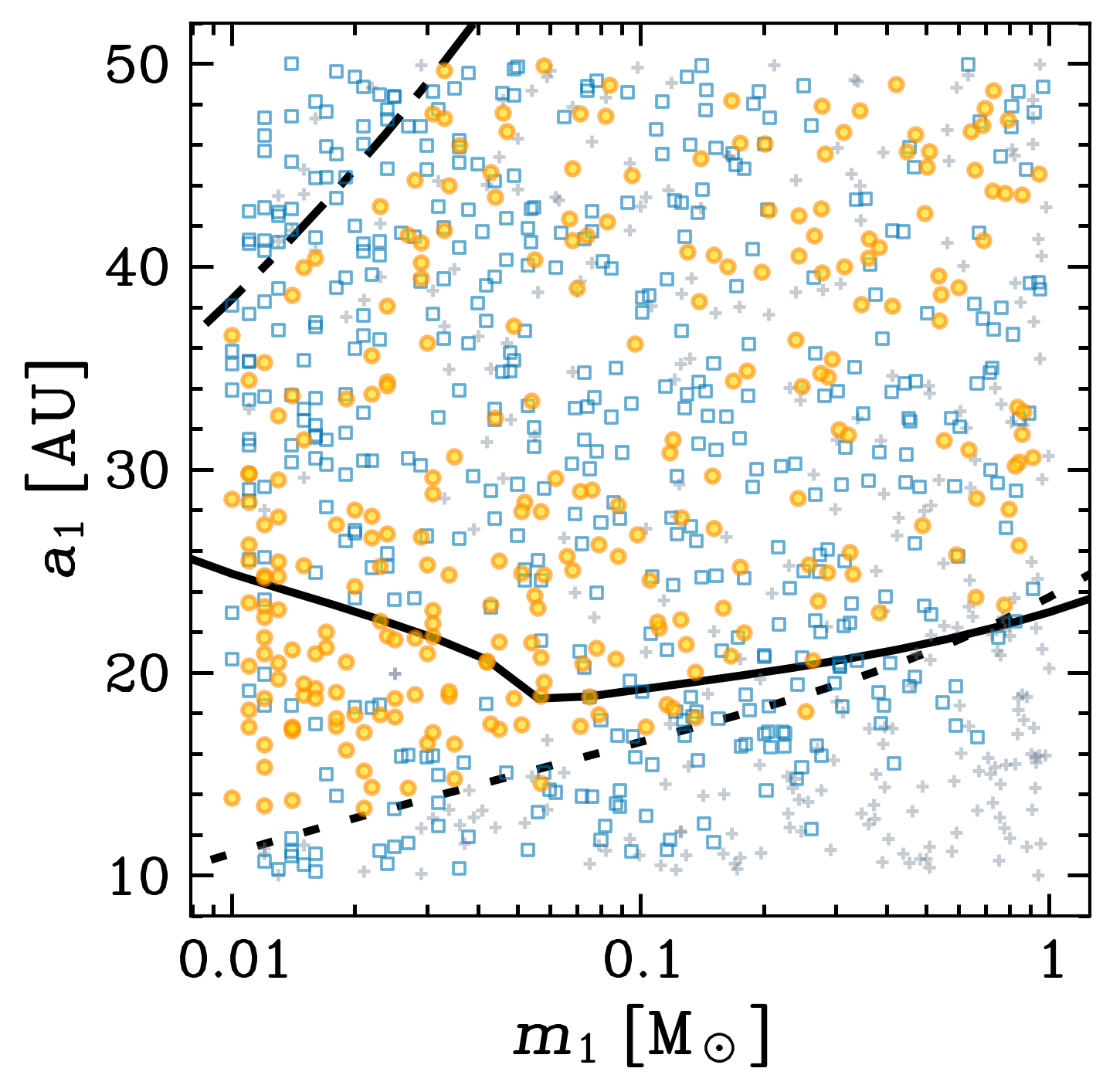}
    \caption{Same as the right panel of Figure~\ref{fig:HAT-P-7_param_space}, but with an initial mutual inclination of $i_{12,0}=60^\circ$ and integration time extended to 400~Myr.}
    \label{fig:HAT-P-7_60inc}
\end{figure}

Above, we focused on the idealized case where $i_{12,0} = 90^\circ$.
For the case of more general orientations, the evolution is not as straightforward to evaluate precisely -- again, due to the chaotic octupole-order evolution.
To begin, we repeat the \textit{N}-body simulations above with a more representative initial condition of $i_{12,0} = 60^\circ$\footnote{Values of $i_{12, 0}$ spanning roughly $55^\circ$--$125^\circ$ yield comparable outcomes \citep{Yang2025}, consistent with the expected range of the octupole-active ZLK window for this architecture \citep[e.g.][]{munoz2016_octwindow}.}.
In this case, the inner companion is still expected to reach $e_{1,\lim}$, but only over octupole-order ZLK timescales (see Section~\ref{ssec:timescale}) and in a chaotic manner.
The integrations are extended to 400~Myr to accommodate the longer timescales.
The resulting outcomes are shown in Figure~\ref{fig:HAT-P-7_60inc}.

In this more realistic setup, the overall behavior remains qualitatively similar, but with several notable differences. First, a subset of systems in Region (2) failed to form HJs within the integration time. This outcome reflects the chaotic nature of octupole-level ZLK cycles, which produce a long tail in the distribution of timescales required to reach $e_{1,\max}$.
Second, the fraction of systems that successfully form HJs -- relative to those that eject or tidally disrupt their planets -- increases across most regions of parameter space when compared to the more extreme initial conditions shown in Figure~\ref{fig:HAT-P-7_param_space}. 

This second trend highlights a key feature of the EC: for most initial inclinations ($i_{12,0} < 90^\circ$), the inner companion's eccentricity evolves gradually toward the instability boundary, rather than crossing it abruptly. Consequently, the inner system stays on the \textit{verge} of instability, efficiently exciting planetary eccentricities while keeping the likelihood of immediate ejection low. Unlike standard three-body dynamical instability where ejection dominates, the EC produces comparable fractions of tidal disruptions and ejections (see Figure~8 of \citealt{Yang2025}).

\subsection{Characteristic timescales}\label{ssec:timescale}
We now estimate the timescale for the EC. During each ZLK cycle, the fraction of time that the inner companion spends near $e_1 \approx e_{1,\max}$ is roughly approximated by \citep[e.g.][]{Liu2015}:
\begin{equation}
    \frac{t_{\rm peak}}{t_{\rm ZLK}} \sim \sqrt{1-e_{1,\max}^2},
    \label{eq:t_peak}
\end{equation}
where the ZLK timescale is given by Equation~\eqref{eq:t_ZLK}.
Then, if the inner companion consistently attains some constant maximum eccentricity $e_{1, \max} > e_{1, \rm crit}$ over successive ZLK cycles,
the time required to excite the planet's eccentricity to an extreme value is set by the instability timescale modulated by the duty cycle of high-e phases:
\begin{equation}
    t_{\rm EC} = \frac{t_{\rm instab}}{t_{\rm peak}/t_{\rm ZLK}} \sim \frac{10^3 (K-K_{\rm crit})^{-1.1}}{(1-e_{1,\max}^2)^{1/2}}P_1.
    \label{eq:t_ec}
\end{equation}
When octupole-order effects are important, the maximum eccentricity attained by the inner companion over successive ZLK cycles varies.
Nevertheless, we can still evaluate Equation~\eqref{eq:t_ec} with $e_{1, \max}$ equal to the largest attained eccentricity over multiple ZLK cycles.
Since this maximum value is not attained every ZLK cycle, this procedure yields a \emph{lower bound} on the EC timescale.

We now apply the results to the HAT-P-7 system, incorporating observational constraints on the orbit of the inner companion. Specifically, we adopt its mass, semimajor axis, and eccentricity to be $0.21\rm M_\odot$, 28~au, and 0.5, respectively, based on the setup in \citet{Yang2025}. We then randomly sample the mutual inclination between the inner and outer companions from the interval $[40^\circ,140^\circ]$ and ran simulations to 400~Myr. For reference, the estimated system's age is $2$~Gyr \citep{Bonomo2017}, the quadrupole-order ZLK timescale is $t_{\rm ZLK} \approx 1.3$~Myr, and the octupole timescale is $t_{\rm EKM} \approx 117$~Myr.

\begin{figure}
    \centering
    \includegraphics[width=0.75\linewidth]{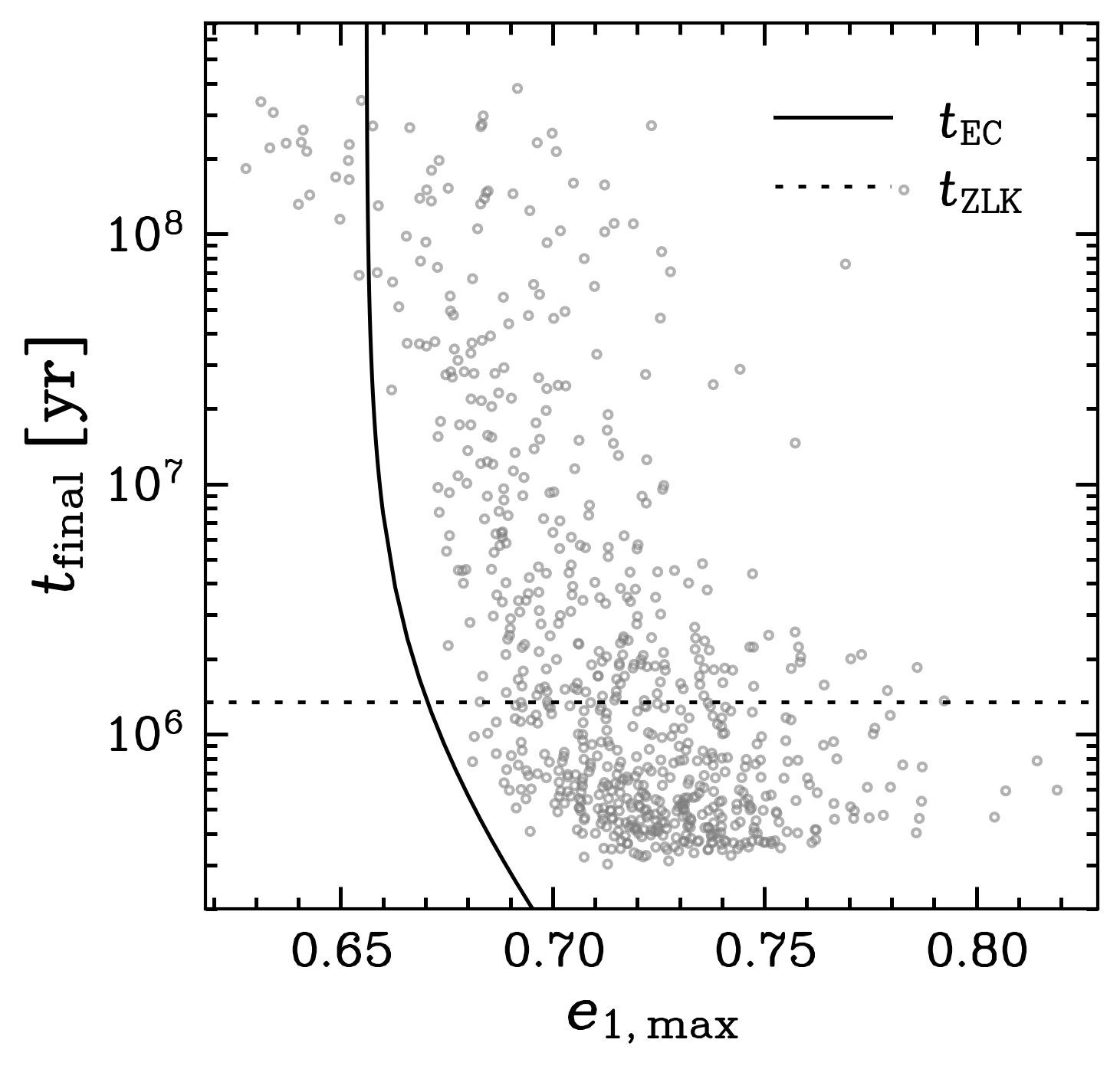}
    \caption{Migration timescale of the planet as a function of the maximum eccentricity reached by the inner companion from numerical simulations based on the observed properties of the HAT-P-7 system. The solid curve shows the duty-cycle-modulated instability time, as given by Equation~\eqref{eq:t_ec}, while the dashed line represents the ZLK timescale from Equation~\eqref{eq:t_ZLK}.}
    \label{fig:HAT-P-7_timescale}
\end{figure}

Figure~\ref{fig:HAT-P-7_timescale} shows the migration time -- defined as the epoch when the planet becomes tidally circularized -- as a function of the maximum eccentricity attained by the inner companion in the simulations. The analytical lower bound on the EC timescale agrees well with the numerical results. For most systems that successfully form HJs, the inner companion reaches $e_{1,\rm crit} = 0.656$ and triggers planet migration within a single ZLK timescale.
A small number of systems with $e_{1,\max} < e_{1,\rm crit}$ still form HJs.
We attribute this to the approximate nature of the diffusion threshold: as shown in Figure~\ref{fig:T_vs_K}, some systems with $K < K_{\rm crit}$ can nevertheless undergo slow orbital diffusion.

\section{Discussion}\label{sec:discussion}

In this section, we discuss our results in the context of existing theoretical and future observational work.

\subsection{Comparison to previous work on orbital stability in three-body systems}\label{ssec:disc_stability}
Our work in Section~\ref{sec:dynamics} considers the stability of S-type orbits, in which a planet orbits one star in a stellar binary.
This problem has been extensively studied, beginning with the seminal work of HW99. Their empirical stability boundary gives the critical semimajor axis beyond which an initially circular planet becomes unstable, as a function of binary separation, eccentricity, and mass ratio. Subsequent numerical studies \citep[e.g.][]{Pilat-Lohinger2002,Quarles2020} have expanded the parameter space and generally confirmed the results in HW99.

Despite the extensive numerical work, the underlying mechanism for instability is not completely understood. 
\citet{Mardling1999} derived a widely used empirical stability boundary based on an analogy with chaotic energy exchange in the binary-tides problem, calibrated primarily for stellar triples with roughly comparable component masses. \citet{Mudryk2006} proposed that instability arises from the overlap of mean-motion resonances between the inner and outer orbits. Each of these models has important limitations: the former one is less reliable when extrapolated to extreme mass ratios (e.g., planets in binary systems), while the latter approach relies on a first-order mass-ratio expansion that fails for massive companions and diverges for binary eccentricities exceeding $\sim 0.66$ (a known limitation of the expansion; see \citealp{Murray1999}).
The iterative map developed in Section~\ref{sec:dynamics} provides a complementary framework for modeling S-type instability. Unlike previous perturbative approaches, it remains well-behaved at high binary eccentricities and large binary mass ratios. It also yields good agreement with the empirical stability boundary from HW99 (see Figure~\ref{fig:stab_boundary}).

We note that our model predicts a smooth stability boundary, whereas \citet{Mudryk2006} identified jagged features associated with resonance overlap. This difference may reflect the distinct dynamical regimes considered, as their analysis focused on lower mass ratios and smaller period ratios.
In such a regime, the kick function may no longer be well-approximated as sinusoidal \citep[e.g.][]{Hadden2024}, in which case the approximation in Section~\ref{ssec:the_map} would no longer be accurate.
High-resolution numerical studies could help identify whether indeed different instability mechanisms operate across different architectures.

Previous studies have also examined the effects of planetary eccentricity and inclination on orbital stability through numerical simulations \citep[e.g.,][]{Pilat-Lohinger2002, Quarles2020}. While our derivation of Equation~\eqref{eq:e_crit} assumes an initially circular and coplanar planet orbit, the underlying formalism is more general. In particular, Equation~\eqref{eq:de_1} provides a full expression for the change in the planet's orbital energy, and hence semimajor axis, resulting from close encounters, explicitly accounting for the planet's eccentricity and inclination. This generalized form could be used to derive analogous diffusion thresholds and stability boundaries for systems with initially inclined or eccentric planetary orbits, providing a theoretical counterpart to the trends observed in those numerical studies.

Our results can also be compared to studies of instability timescales. \citet{David2003} and \citet{Fatuzzo2006}, for example, found that the survival time of an Earth-mass planet in S-type systems scales exponentially with periastron distance. Our analytical instability timescale in Equation~\eqref{eq:t_D} shows qualitatively similar behavior. We refrain from detailed quantitative comparisons, however, since those studies averaged over broad distributions of binary eccentricities.

\begin{figure}
    \centering
    \includegraphics[width=1\linewidth]{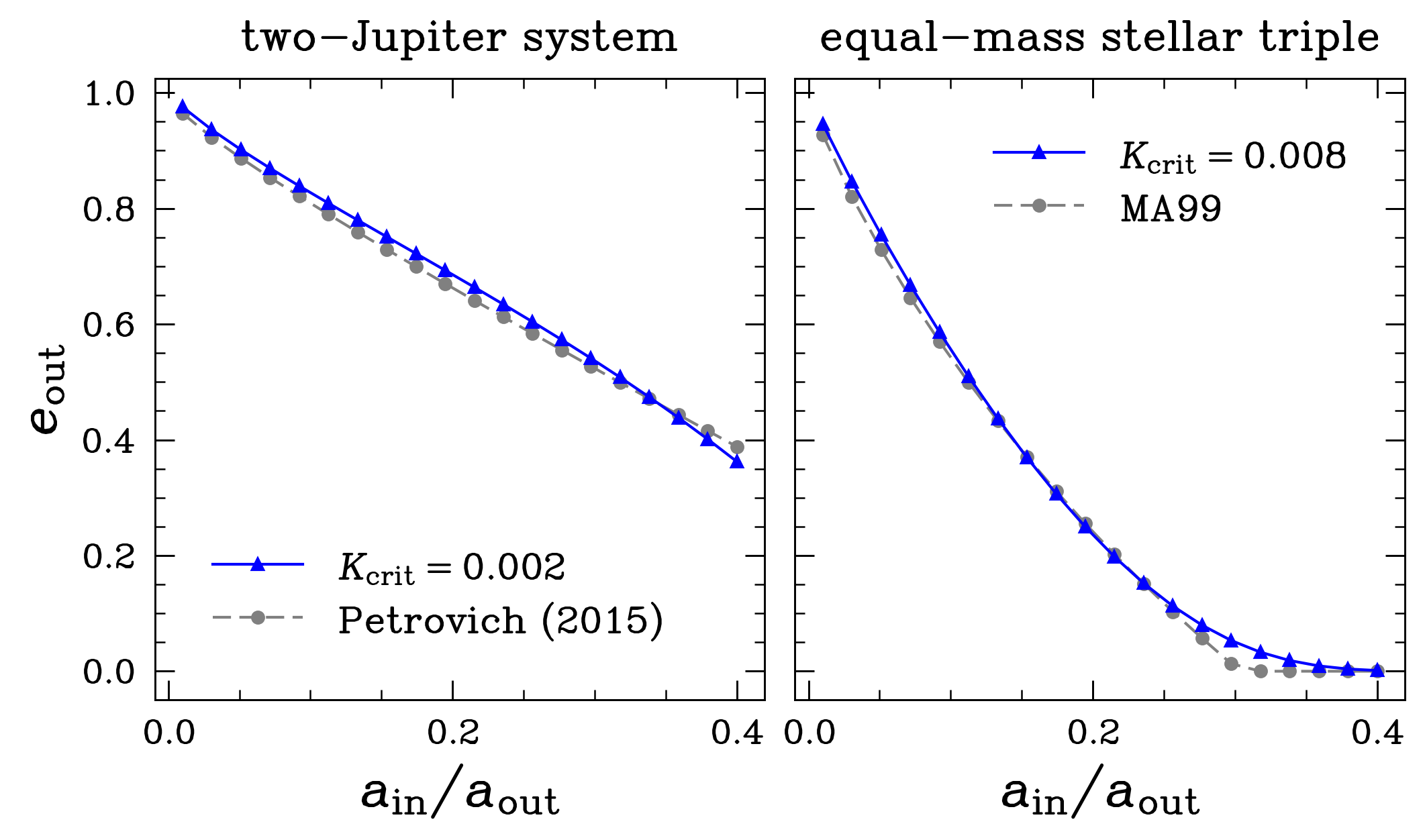}
    \caption{Comparison of our analytical stability criterion (Equation \ref{eq:e_crit}) with empirical boundaries for hierarchical systems. \textbf{Left:} a two-Jupiter system, where setting $K_{\rm crit}=0.002$ reproduces the empirical criteria from \citet{Petrovich2015b}. \textbf{Right:} an equal-mass stellar triple, where $K_{\rm crit}=0.008$ matches the fitted stability boundary from \citet[MA99]{Mardling1999}.}
    \label{fig:compare_other}
\end{figure}

Moreover, by tuning $K_{\rm crit}$ in Equation~\eqref{eq:e_crit}, we can extend the same closed-form expression to qualitatively reproduce the stability boundaries of both hierarchical two-planet and stellar triple systems.
In the left panel of Figure~\ref{fig:compare_other}, we recover the empirical stability boundary for two-planet systems from \citet{Petrovich2015b} for Jupiter-mass planets by setting $K_{\rm crit}=0.002$.
In the right panel, we match the stability boundary for equal-mass stellar triples from \citet{Mardling1999} by adopting $K_{\rm crit}=0.008$. These values of $K_{\rm crit}$ are notably smaller than the $K_{\rm crit} = 0.12$ obtained in Section~\ref{sec:dynamics} for S-type systems, and are even farther away from the $K_{\rm crit} \sim 1$ obtained for the classical Chirikov standard map \citep[e.g.][]{ott2002_chaosdynamicalsystemsbook}.
This trend is consistent with our interpretation that reduced values of $K_{\rm crit}$ arise from the contribution of higher harmonics in the kick function \citep[cf.][]{Hadden2024}: lower-mass outer bodies must pass closer to the middle body to generate comparable kick amplitudes, yielding a more sharply peaked kick function and enhanced resonance overlap.

Despite the differences in $K_{\rm crit}$, the good qualitative agreement across planetary and stellar regimes underscores the general applicability of our analytic stability criterion.
Further exploration of how $K_{\rm crit}$ scales with system mass ratios may help unify stability prescriptions for hierarchical systems across the full range of architectures.

\subsection{Mass dependence in the EC}
Although the EC scenario was originally proposed for the HAT-P-7 system, where both companions are stellar, we have shown that it remains effective even when the inner companion has a mass as low as $10M_J$. This is because the diffusion threshold scales weakly with the mass ratio between the host star and the inner companion (see Equation~\ref{eq:e_crit}).

For lower-mass inner companions, however, a key change arises: as $m_1$ decreases, the coupling between the planet and inner companion weakens, and the coplanarity condition (Equation~\ref{eq:Omega}) may no longer be satisfied. In this regime, the system falls outside the EC domain, and alternative migration channels are expected to dominate (see Region~(1) of Figure~\ref{fig:HAT-P-7_60inc} and Section~\ref{ssec:necessary_criteria}).

Additionally, as the inner companion's mass decreases, closer pericenter approaches are required to produce sufficiently large semimajor axis kicks to trigger diffusive/chaotic evolution.
For closer approaches, the kick function may no longer be well-described as sinusoidal, and a more detailed analysis including all harmonics of the kick function may be necessary to characterize the instability threshold \citep{Hadden2024}.
While the parameter space explored here does not require such refinements, they may become essential for even lower-mass companions than the ones considered in this work.

\subsection{Application to HD 80606 and TIC 241249530}\label{ssec:app}
In the HAT-P-7 system, the discovery of an inner companion alleviated the difficulty of forming the HJ through interactions with the known distant companion alone. Motivated by this example, we now ask what kinds of inner companions would be required to resolve the same fine-tuning problem in two other well-known systems that host HJ progenitors and distant companions: HD 80606 and TIC 241249530.

HD 80606b ($m_p=4.1\,\rm M_J$, $a_p=0.46$~au, $e_p=0.93$; \citealp{Naef2001}) is thought to be undergoing high-e migration, potentially via the ZLK mechanism driven by its main-sequence companion at a projected separation of $\sim 1000$~au. However, this requires a fine-tuned initial mutual inclination of $\sim 85^\circ$--$95^\circ$ between the planet and the distant companion \citep{Wu2003}. 

Similarly, TIC 241249530 hosts a HJ progenitor ($m_p=5.0$~$\rm M_J$, $a_p=0.64$~au, $e_p=0.94$; \citealp{Gupta2024}) and a low-mass stellar companion, TIC 241249532, at a projected separation of 1664~au. \cite{Gupta2024} showed that the present-day parameters of the system can be reproduced if the planet initially formed beyond 7~au and had a mutual inclination of $86.8^\circ$--$93.2^\circ$ with respect to the companion.

\begin{figure}
    \centering
    \includegraphics[width=1\linewidth]{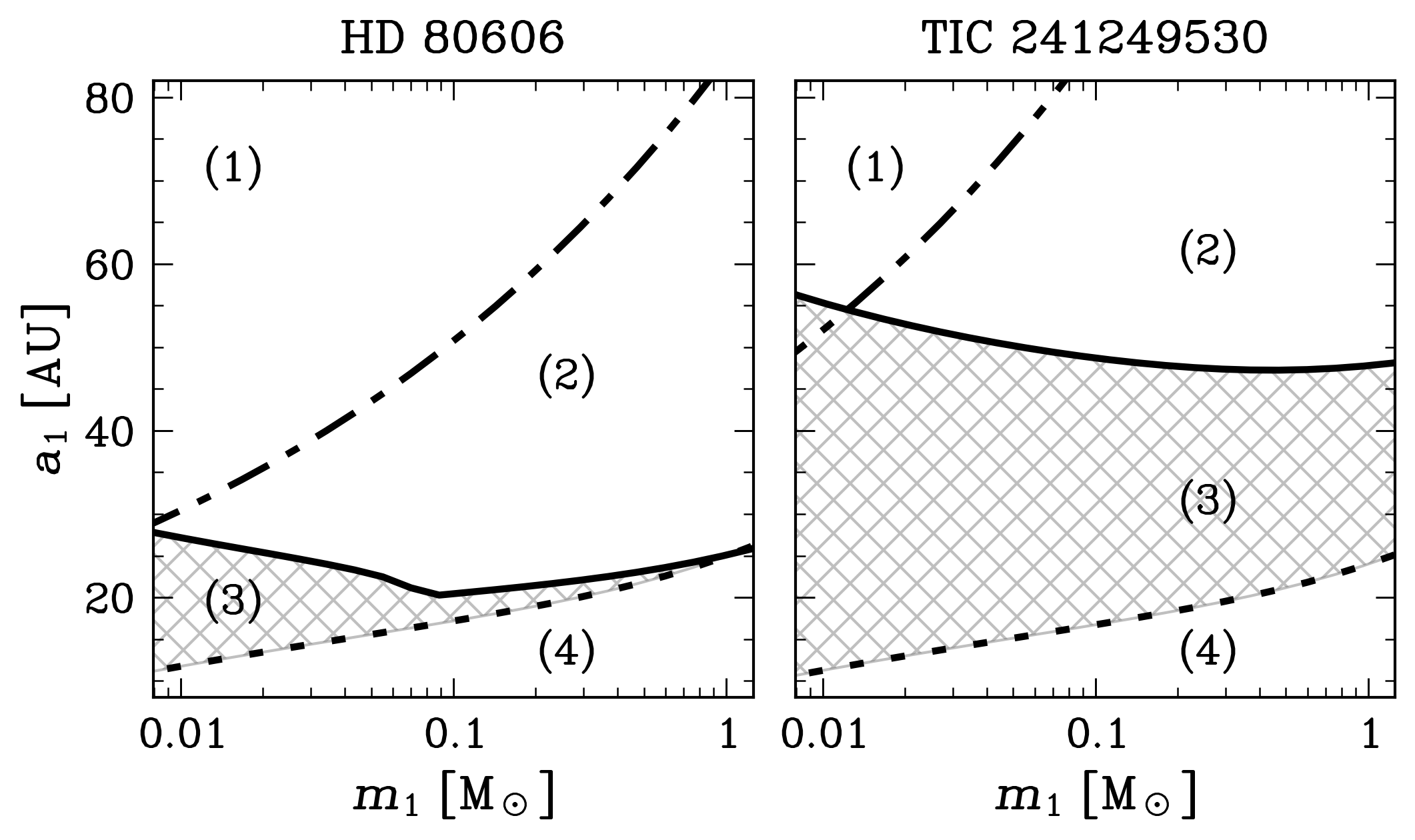}
    \caption{Same as the left panel of Figure~\ref{fig:HAT-P-7_param_space}, but applied to the HD 80606 and TIC 241249530 systems. The shaded regions indicate where the EC is predicted to be strongly suppressed, for the system parameters listed in Table~\ref{tab:param_HD_TIC}.}
    \label{fig:other_systems}
\end{figure}

In both systems, the narrow range of mutual inclinations that are required in the standard ZLK scenario motivate searches for additional, closer-in companions that could facilitate a four-body migration pathway, such as the EC, thereby relaxing the fine-tuning needed in standard ZLK scenarios.
Figure~\ref{fig:other_systems} illustrates the parameter space in which a hypothetical inner companion could enable the EC in each of the systems. The orbital parameters used in these calculations are listed in Table~\ref{tab:param_HD_TIC}, based primarily on values adopted by \citet{Wu2003} and \citet{Gupta2024}.

\begin{table}[t]
    \centering
    \caption{Parameters used in the calculations for HD 80606 and TIC 241249530.}
    \begin{threeparttable}
        \begin{tabularx}{\columnwidth}{l l}
            \hline\hline
            Parameters & Values\\
            \hline\hline
            HD 80606 & \\
            \hline
            $M_\star$, $m_p$, $m_2$ & 1.05~$\rm M_\odot$, 4.38~$\rm M_J$, 1.10~$\rm M_\odot$\\
            $a_p$, $a_2$ & 3~au, 1000~au\\
            $e_p$, $e_2$ & 0, 0.5\\
            \hline
            TIC 241249530 & \\
            \hline
            $M_\star$, $m_p$, $m_2$ & 1.27~$\rm M_\odot$, 4.98~$\rm M_J$, 0.45~$\rm M_\odot$\\
            $a_p$, $a_2$ & 3~au, 1664~au\\
            $e_p$, $e_2$ & 0, 0.5\\
            \hline
        \end{tabularx}
        
    \end{threeparttable}
    \label{tab:param_HD_TIC}
\end{table}

\subsection{Systems where chaotic diffusion may drive HJ migration}
While the previous section focused on predicting inner companions in hot Jupiter systems with distant stellar perturbers, our analytical framework for chaotic diffusion may also apply to systems with just a single eccentric companion. 
Notably, WASP-53b and WASP-81b are hot Jupiters with brown dwarf companions on eccentric, close-in orbits: WASP-53c ($m\sin{i}>16.4~\rm M_J$, $a>3.73$~au, $e = 0.84$; \citealp{Triaud2017}) and WASP-81c ($m\sin{i}=56.6~\rm M_J$, $a=2.43$~au, $e=0.56$; \citealp{Triaud2017}). The present-day compactness of these systems suggests that they were even more tightly packed during the proto-hot Jupiter phase.
If the planet and companion are significantly misaligned, migration may have been driven by mechanisms such as planet–planet ZLK \citep[e.g.,][]{Naoz2011} or dynamical scattering \citep[e.g.,][]{Petrovich2014, Lu2025}. However, if the system is close to coplanar, the planet could still migrate under perturbations from the brown dwarf either through chaotic diffusion (as described in Section~\ref{sec:dynamics}) or via coplanar high-e migration (CHEM; \citealp{Petrovich2015a}). 

CHEM operates through secular forcing from the companion to excite the planet's eccentricity to high values, and its efficiency depends sensitively on the initial conditions of the proto-HJ. In comparison, chaotic diffusion becomes effective when the companion's approach is sufficiently close to the planet that non-secular interactions become important, and it allows the planet to explore a broader region of eccentricity and angular momentum phase space than is typically accessible through secular interactions alone. While CHEM is more sensitive to the planet's initial configuration, diffusion depends more strongly on the companion's properties, and the two mechanisms are likely complementary; both may play a role in HJ formation.

\subsection{Feasibility of the EC: companion statistics}

To form an HJ, the EC requires two additional bodies other than the host star.
While HAT-P-7 remains the only HJ system in which two stellar companions have been observed with properties that satisfy the conditions for the EC \citep{Yang2025}, existing observational studies offer some guidance on how common such configurations may be\footnote{
The GQ Lupi system is a potential progenitor to a HAT-P-7-like architecture, hosting a circumstellar disk, a close $\sim 30 \; \rm M_{\rm Jup}$ companion at $\sim 100\;\mathrm{au}$ on a mildly eccentric orbit, and a wide tertiary companion with a projected separation of $\sim 2400\;\mathrm{au}$ \citep{neuhauser2006_gqlupb, lazzoni2020_gqlupc, alcala2020_gqlupc, venkatesan2025_gqlupb}.}.

Below, we summarize key observational results relevant to the occurrence of such companions and use them to estimate the prevalence of the EC in HJ formation.

First, the EC requires a distant outer stellar-mass companion.
While roughly half of planet-hosting stars have stellar companions \citep[from Kepler and K2 samples; e.g.,][]{horch2014_companions, matson2018_companions}, only about one third possess separations $\gtrsim 10^3$~au \citep{ziegler2021_soartess2}, as considered here.
This fraction may be somewhat higher among giant-planet hosts: \citet{Ngo2016} found that $47^{+7}_{-7}\%$ of HJ systems harbor stellar companions with semimajor axes between 50-–200~au, approximately three times the rate for field stars.
They interpreted this enhancement as tentative evidence that stellar companions may enhance giant-planet formation.
Although most of these companions are too low in mass to drive standard ZLK migration, they could still play a significant role in facilitating HJ formation through the EC mechanism.

Second, the EC requires a relatively massive inner companion.
In HAT-P-7, for example, the inner system hosts an $\sim0.2~{\rm M_\odot}$ M dwarf at $\sim30$~au.
Close stellar companions of this kind are generally thought to inhibit planet formation by truncating the protoplanetary disk \citep[e.g.,][]{Artymowicz1994}, and such suppression is well supported by observations \citep[e.g.,][]{Wang2014, kraus2016_ruinous, Ziegler2020}.
Combining results from multiple observational surveys, \citet{moekratter2021_binarystarsplanets} find that planet-hosting stars are 2--5 times less likely to have stellar companions with semimajor axes of 10-–50~au compared to field stars.
This deficit is consistent with that observed among HJ hosts: \citet{Ngo2016} report that only $3.9^{+4.5}_{-2.0}\%$ of HJ hosts harbor stellar companions with masses $\ge 0.08,{\rm M\odot}$ and separations of 1–-50~au, roughly five times lower than for field stars.

However, as shown in this work (e.g., Fig.~\ref{fig:HAT-P-7_60inc}), substellar companions can also efficiently induce the EC.
Analysis of archival GPIES direct imaging data by \citet{Squicciarini2025} reports an occurrence rate of $2.3_{-0.8}^{+1.0}\%$ for companions with $5~{\rm M_J} < m < 80~{\rm M_J}$ at separations of $10~{\rm au}< a < 100~{\rm au}$ around young stars (consistent with earlier estimates, e.g.\ \citealp{kraus2008_bddesert, dieterich2012_bd}), although none of the detected systems are known HJ hosts.
At still lower masses, reanalysis of the California Legacy Survey radial velocity data by \citet{Zink2023} finds that HJ hosts have on average $1.3^{+1.0}_{-0.6}$ giant companions per system, with orbital periods up to 40000 days and typical masses about three times larger than those of the inner HJs.
Together, these results imply that planetary- to substellar-mass companions are relatively common in HJ systems.
Although our analysis does not explicitly model such low-mass perturbers, the results presented here suggest that they may be capable of driving EC-like evolution (see Fig.~\ref{fig:HAT-P-7_60inc}).

While current uncertainties preclude a precise determination of the EC's overall contribution to HJ formation, we can make some coarse estimates.
The conditional probability that an HJ forms via the EC can be expressed as
\begin{equation}
    P_{\rm EC}({\rm HJ})
        =
             P({\rm CJ})
             \underbrace{P({\rm IC | CJ}) P({\rm OC | CJ}) \eta_{\rm mig}}_{\eta_{\rm EC}},
             \label{eq:bayes_PHJ_EC}
\end{equation}
where $P(\rm CJ)$ is the occurrence rate of cold Jupiters, $P(\rm IC | CJ)$ and $P(\rm OC | CJ)$ are the probabilities that a cold Jupiter host also possesses inner and outer companions, respectively. The factor $\eta_{\rm mig} \simeq 0.5$ represents the probability that a system meeting these criteria successfully undergoes migration via the EC (that is, has a favorable mutual inclination and avoids ejection or tidal disruption during the process).
The combined term $\eta_{\rm EC}$ therefore quantifies the overall efficiency of the EC in producing HJs given the presence of cold Jupiters.

Since $P(\rm OC|CJ)$ is not well constrained, we adopt the aforementioned occurrence rate of wide stellar companions to general planetary systems, giving $P(\rm OC|CJ)=1/6$.
A direct estimate from the intrinsic stellar wide-binary distribution expectedly yields a comparable value, as wide binaries do not affect the occurrence rates of planetary systems \citep{moekratter2021_binarystarsplanets}.

For $P{\rm (IC|CJ)} = P{\rm (CJ|IC)} P{\rm (IC)}/P{\rm (CJ)}$, we first consider stellar-mass inner companions.
The intrinsic occurrence rate of such companions is high, $P(\rm IC)\simeq20\%$ over separations of $1$–$50$~au \citep[e.g.,][]{moekratter2021_binarystarsplanets}.
However, close binaries suppress planet formation by a factor of a few, so we adopt $P(\rm CJ|IC)\simeq3\%$, roughly one-fifth of the field cold-Jupiter occurrence rate of $15\%$ \citep{fulton2021_cjrate}.
Substituting into Equation~\eqref{eq:bayes_PHJ_EC} yields $P_{\rm EC}(\rm  HJ)\simeq0.05\%$, about one-tenth of the bias-corrected HJ occurrence rate of $0.5\%$ \citep[e.g.,][]{howard2012_hjoccurrence, wright2012_hjoccurrence, beleznay2022_hjoccurrence, moekratter2021_binarystarsplanets}.

As noted earlier, only $\sim5\%$ of HJ hosts are observed to have stellar-mass companions within $\lesssim50$~au, implying that either many inner companions remain undetected or our assumed $P(\rm  CJ|IC)$ is too high.
In either case, the efficiency of the stellar-mass EC channel appears intrinsically low, consistent with the rarity of close stellar companions among HJ systems.

We next consider the case where the inner companion is a substellar-mass object.
Given the low intrinsic occurrence rate of brown dwarfs at tens of au ($\sim2\%$, as discussed above), even adopting $P(\rm CJ|IC)=1$ yields a hot Jupiter formation rate via the EC of only $\sim0.08\%$.
Using more realistic values of $P(\rm CJ|IC)$ further reduces this estimate, indicating that brown-dwarf inner companions contribute negligibly to EC-driven HJ formation, primarily due to their rarity.

Our estimates remain highly uncertain, particularly because they do not include planetary-mass inner companions ($m_1 \lesssim 10~M_{\rm Jup}$).
This regime lies beyond the analytical scope of this work, as the approximations developed in Section~\ref{sec:dynamics} assume higher-mass perturbers.
Nonetheless, such low-mass companions are far more common than brown dwarfs or stellar companions, and their inclusion could substantially increase the overall EC efficiency.
Future numerical and observational efforts will be required to explore this wider parameter space.

\section{Conclusion}
The eccentricity cascade (EC) was proposed by \citet{Yang2025} as a viable pathway for hot Jupiter (HJ) migration in the HAT-P-7 system, which hosts two stellar companions in a strongly hierarchical configuration. This mechanism suggests that stellar companions traditionally considered too distant or too low in mass to induce migration may nonetheless facilitate HJ formation when aided by an intermediate perturber. In this work, we have identified the underlying mechanism behind the EC and derived analytical criteria that delineate the parameter space in which it operates.

The core of the EC lies in the chaotic diffusion of the planet's eccentricity, driven by perturbations from the inner companion as the system approaches the threshold for S-type orbital instability. In Section~\ref{sec:dynamics}, we analyzed a simplified three-body configuration consisting of the host star, the planet, and an inner companion with fixed eccentricity. Within this framework, we derived closed-form expressions for the stability boundary of S-type orbits and the characteristic instability timescale (see Equations~\ref{eq:e_crit}--\ref{eq:t_D}; Figures~\ref{fig:T_vs_K}--\ref{fig:stab_boundary}).
In Section~\ref{sec:criteria}, we applied this model to a hierarchical four-body system, where an outer stellar companion excites the inner companion's eccentricity through ZLK oscillations.
We derived the necessary condition for the onset of the EC and mapped the parameter space in which the mechanism operates (see Equation~\ref{eq:criteria}, Figure~\ref{fig:HAT-P-7_param_space}).
Our results suggest that, even in binaries where the secondary is distant and/or has low mass, HJ formation can still be facilitated via the EC for a wide range of intermediate companions.
Further observations are required to better characterize the prevalence and importance of the EC towards understanding the origin of HJs.

\begin{acknowledgements}
    We thank Joshua Winn for fruitful discussions and invaluable feedback throughout the course of this work. We are also grateful to Chris Hamilton, Cristobal Petrovich, Caleb Lammers, Tiger Lu, Smadar Naoz, and Yanqin Wu for their helpful comments.
    YS acknowledges support by the Lyman Spitzer Jr.\ Postdoctoral Fellowship at Princeton University and by the Natural Sciences and Engineering Research Council of Canada (NSERC) [funding reference CITA 490888-16].
\end{acknowledgements}

\software{astropy \citep{astropy}, rebound \citep{rebound}, smplotlib \citep{smplotlib}}

\appendix
\section{Derivation for \texorpdfstring{$\delta a_{p,\max}$}{delta ap max} \label{app:delta_a}}

We seek to evaluate the change to the semi-major axis of a planet orbiting a host star when perturbed by an external companion.
Our general approach is inspired by \citet{Heggie2006} but necessarily differs due to the consideration of an elliptical rather than hyperbolic companion.

The Keplerian orbits of the planet and the perturber are described by
\begin{align}
    &\mathbf{r} = a(\cos{E}-e)\, \hb{a} + a \sqrt{1-e^2}\sin{E}\,\hb{b},\\
    &\mathbf{R} = a'(\cos{E'}-e')\, \hb{A} + a'\sqrt{1-e'^2}\sin{E'}\,\hb{B},\label{eq:R}\\
    &R=a'(1-e'\cos{E'}), \quad n't = E'-e'\sin{E'} \label{eq:nt}
\end{align}
where $a$, $e$ and $E$ are the semimajor axis, the eccentricity and the eccentric anomaly of the planet, and $\hb{a}$, $\hb{b}$ are unit vectors aligned with the axes of its orbit. The quantities $a'$, $e'$, $E'$, $\hb{A}$, and $\hb{B}$ are the corresponding orbital elements and basis vectors for the perturber.

The change in the planet's semimajor axis is directly related to the change in its specific orbital energy $\delta\epsilon$ via
\begin{equation}
    \frac{\delta{a}}{a} = -\frac{\delta \epsilon}{\epsilon}.
    \label{eq:a_and_epsilon}
\end{equation}
From first-order perturbation theory, the change in energy due to the external perturber is
\begin{equation}
    \delta\epsilon = -\frac{Gm_1m_2m_3}{M_{12}} \int \dot{\mathbf{R}}\cdot\frac{\partial \mathcal{R}}{\partial\mathbf{R}}\,dt
    = -\frac{Gm_1m_2m_3}{M_{12}}\Int \frac{d\mathbf{R}}{dE'}\cdot\frac{\partial \mathcal{R}}{\partial\mathbf{R}}\,dE', \label{eq:delta_epsilon}
\end{equation}
where $m_1$, $m_2$, $m_3$ are the masses of the host star, the planet and the perturber, and $M_{12}=m_1+m_2$. 

Following the formulation in \citep{Roy2003}, the disturbing function at quadrupole order is
\begin{equation}
    \mathcal{R} = \frac{1}{R^5} \left\{\left[\frac{3}{2}b_1a^2(\hb{a}\cdot\mathbf{R})^2 - \frac{3}{2}b_2a^2(1-e^2)(\hb{b}\cdot\mathbf{R})^2 - \frac{1}{2}eb_3a^2R^2\right]\cos{M} + 3b_4a^2\sqrt{1-e^2}(\hb{a}\cdot\mathbf{R})(\hb{b}\cdot\mathbf{R})\sin{M}\right\},
    \label{eq:disturbing_func}
\end{equation}
where $M=n(t-t_0)$ is the mean anomaly of the planet. The coefficients are Bessel functions of the planet's eccentricity:
\begin{equation}
\begin{aligned}
    &b_1 = J_{-1}(e) - 2eJ_0(e) + 2eJ_2(e) - J_3(e),\quad b_2 = J_{-1}(e) - J_3(e),\\
    &b_3 = eJ_{-1}(e)-2J_0(e) + 2J_2(e)-eJ_3(e), \quad b_4 = J_{-1}(e)-eJ_0(e)-eJ_2(e) + J_3(e).
    \label{eq:Bessel}
\end{aligned}
\end{equation}

Substituting Equations~\eqref{eq:R} and \eqref{eq:disturbing_func} into Equation~\eqref{eq:delta_epsilon}, we obtain the integrand in the form:
\begin{equation}
\begin{aligned}
    \frac{d\mathbf{R}}{dE'} \cdot \frac{\partial \mathcal{R}}{\partial\mathbf{R}} = &~ \frac{3}{2}c_1a^2 \left(\frac{2a'^2f_1(E')}{R^5} - \frac{5a'^3e'f_2(E')}{R^6}\right)\cos{M} - \frac{3}{2}c_2a^2 \left(\frac{2a'^2\tilde{f}_1(E')}{R^5} - \frac{5a'^3e'\tilde{f}_2(E')}{R^6}\right)\cos{M} \\
    &~ + \frac{3}{2}c_3a^2 \left(\frac{a'e'\sin{E'}}{R^4}\right) \cos{M}  + 3c_4a^2 \left(\frac{a'^2 f_3(E')}{R^5} + \frac{a'^2 \tilde{f}_3(E')}{R^5} - \frac{5a'^3e'f_4(E')}{R^6}\right) \sin{M},
    \label{eq:integrand}
\end{aligned}
\end{equation}
where we have redefined the Bessel-related coefficients in Equation~\eqref{eq:Bessel} as
\begin{equation}
    c_1 \equiv b_1, \quad
    c_2 \equiv b_2(1-e^2), \quad
    c_3 \equiv b_3e, \quad
    c_4 \equiv b_4\sqrt{1-e^2}.
    \label{eq:rescale_Bessel}
\end{equation}
The functions $f_i(E')$ and $\tilde{f}_i(E')$ encapsulate the geometric dependence of the perturbation on the position of the external companion and are given by:
\begin{align}
    f_1(E') =\;& -\sin{E'}(\cos{E'}-e')(\hb{a}\cdot\hb{A})^2 + (1-e'^2)\sin{E'}\cos{E'} (\hb{a}\cdot\hb{B})^2 \notag \\
            & + \sqrt{1-e'^2} (\cos{E'}(\cos{E'}-e')-\sin^2{E'}) (\hb{a}\cdot\hb{A})(\hb{a}\cdot\hb{B}), \label{eq:f1}\\
    f_2(E') =\;& \sin{E'}(\cos{E'}-e')^2(\hb{a}\cdot\hb{A})^2 + (1-e'^2)\sin^3{E'} (\hb{a}\cdot\hb{B})^2 \notag \\
            & + 2\sqrt{1-e'^2} \sin^2{E'}(\cos{E'}-e') (\hb{a}\cdot\hb{A})(\hb{a}\cdot\hb{B}),\\
    f_3(E') =\;& -\sin{E'}(\cos{E'}-e')(\hb{a}\cdot\hb{A})(\hb{b}\cdot\hb{A}) - \sqrt{1-e'^2}\sin^2{E'} (\hb{a}\cdot\hb{A})(\hb{b}\cdot\hb{B}) \notag\\
    & +\sqrt{1-e'^2}\cos{E'}(\cos{E'}-e')(\hb{a}\cdot\hb{B})(\hb{b}\cdot\hb{A})
    + (1-e'^2)\sin{E'}\cos{E'} (\hb{a}\cdot\hb{B})(\hb{b}\cdot\hb{B}),\\
    f_4(E') =\;& \sin{E'}(\cos{E'}-e')^2 (\hb{a}\cdot\hb{A})(\hb{b}\cdot\hb{A}) + (1-e'^2)\sin^3{E'}(\hb{a}\cdot\hb{B})(\hb{b}\cdot\hb{B}) \notag\\
    & + \sqrt{1-e'^2}\sin^2{E'}(\cos{E'}-e') [(\hb{a}\cdot\hb{A})(\hb{b}\cdot\hb{B})+(\hb{a}\cdot\hb{B})(\hb{b}\cdot\hb{A})],
    \label{eq:f4}
\end{align}
The corresponding functions $\tilde{f}_i(E')$ are obtained by exchanging $\hb{a} \leftrightarrow \hb{b}$ in $f_i(E')$.
Given Equation~\eqref{eq:nt}, the planet's mean anomaly can be expressed as $\cos{M}=\Re [\Exp{in(t-t_0)}] = \Re[\Exp{i\frac{n}{n'}(E'-e'\sin{E'})-int_0}]$. Substituting this into Equation~\ref{eq:integrand}, the integral for $\delta\epsilon$ can be written as the sum of many terms of form
\begin{equation}
    \Re \left\{ \Exp{-int_0}\Int \frac{f(E')}{R^k} \Exp{i\frac{n}{n'}(E'-e'\sin{E'})} \, dE'\right\},\label{eq:app_deltaeps_termform}
\end{equation}
for some integer $k$ and some function $f(E')$ composed of terms defined in Equations~\eqref{eq:f1}-\eqref{eq:f4}.
To evaluate such integrals, we introduce the deceptively natural result (derived in Appendix~\ref{sec:contourintegrals})
\begin{equation}
    \Int \frac{f(E')}{R^k}\Exp{\F}\,dE'
    \approx 2\pi i\,{\rm Res}\left[\frac{f(E')}{R^k}\Exp{\F}; {E'_0}\right],
    \label{eq:res}
\end{equation}
where $E'_0 = \cos^{-1}(1/e')$ is the complex value of $E'$ where $R = 0$, Res denotes the complex residue \citep[e.g.][]{brownchurchill}, and we've defined the phase function
\begin{equation}
    \F\equiv i\frac{n}{n'}(E'-e'\sin{E'}).
\end{equation}
Using this result, we find
\begin{equation}
\begin{aligned}
\delta{\epsilon} \approx
    &~ -6\pi \frac{Gm_1m_2m_3}{M_{12}}a^2a'^2\, \Re \Biggl\{ i \Exp{-int_0} \times \Exp{-\frac{n}{n'}\left(\cosh^{-1}(1/e')-\sqrt{1-e'^2}\right)} \times\\
    &~ \biggr[\frac{e'}{2a'}c_3 {\xi_0}
     + \left(-c_1(\hb{a}\cdot\hb{A})^2 + c_2(\hb{b}\cdot\hb{A})^2 + 2ic_4(\hb{a}\cdot\hb{A})(\hb{b}\cdot\hb{A})\right) {\xi_1} \\
    &~~ + \left(c_1(\hb{a}\cdot\hb{B})^2 - c_2(\hb{b}\cdot\hb{B})^2 -2ic_4(\hb{a}\cdot\hb{B})(\hb{b}\cdot\hb{B})\right) {\xi_2} \\
    &~~ + \left(c_1 (\hb{a}\cdot\hb{A})(\hb{a}\cdot\hb{B}) - c_2 (\hb{b}\cdot\hb{A})(\hb{b}\cdot\hb{B}) - ic_4[(\hb{a}\cdot\hb{B})(\hb{b}\cdot\hb{A})+(\hb{b}\cdot\hb{B})(\hb{a}\cdot\hb{A})]\right) {\xi_3}
    \biggr]\Biggl\},
    \label{eq:de_1}
\end{aligned}
\end{equation}
where
\begin{equation}
\begin{aligned}
    \xi_0 = \frac{n/n'}{3a'^4e'(1-e'^2)^{3/2}}, \quad
    \xi_1 = \frac{n/n'}{4a'^5(1-e'^2)^{3/2}} - \frac{(n/n')^2}{12a'^5e'^2}, \quad
    \xi_2 = -\frac{n/n'}{4a'^5(1-e'^2)^{3/2}} - \frac{(n/n')^2}{12a'^5e'^2}, \quad
    \xi_3 = \frac{i(n/n')^2}{6a'^5e'^2}.
\end{aligned}
\end{equation}

\subsection{Circular and coplanar limit}
For $e\lesssim0.5$, the Bessel-related coefficients in Equation~\eqref{eq:rescale_Bessel} scales linearly with $e$.
Specifically,
\begin{equation}
    c_1 = -\frac{5}{2}e, \quad c_2=-\frac{1}{2}e, \quad c_3=-2e, \quad c_4=-\frac{3}{2}e. \label{eq:Bessel_value}
\end{equation}
In the case where the planet and the perturber are coplanar, we can choose $\hb{a}=(1,0)$ and $\hb{b}=(0,1)$. Then
\begin{equation}
    \hb{A} = (\cos{\varpi}, \sin{\varpi}), \quad \hb{B}=(-\sin{\varpi}, 
    \cos{\varpi}), \label{eq:coord}
\end{equation}
where $\varpi=\Omega + \omega$ of the perturber.
With these choices, Equation~\eqref{eq:de_1} simplifies to
\begin{equation}
\begin{aligned}
    \delta\epsilon
    =&-2\pi e \frac{Gm_1m_2m_3}{M_{12}} \frac{a^2}{a'^3}\left[\frac{n/n'}{(1-e'^2)^{3/2}}\sin{nt_0} - \frac{3(n/n')^2}{e'^2} \sin(2\varpi+nt_0)\right] \Exp{-\frac{n}{n'}\left(\cosh^{-1}(1/e')-\sqrt{1-e'^2}\right)}.
\end{aligned}
\end{equation}
The resulting maximum $\delta{a}$ is then given by Equation~\eqref{eq:a_and_epsilon}:
\begin{equation}
    \frac{\delta a_{\max}}{a} = -\frac{\delta\epsilon_{\max}}{\epsilon} = 2\pi e\frac{m_3 a^3}{M_{12} a'^3} \left[\frac{n/n'}{(1-e'^2)^{3/2}} +\frac{3(n/n')^2}{e'^2}\right] \Exp{-\frac{n}{n'}\left(\cosh^{-1}(1/e')-\sqrt{1-e'^2}\right)}.
    \label{eq:delta_a_max}
\end{equation}

Note that $\delta a_{\max}$ in Equation~\eqref{eq:delta_a_max} formally vanishes for a planet on a circular orbit ($e=0$). Nonetheless, the planet can still acquire eccentricity due to the perturbation from an external companion. At quadrupole order, the instantaneous change in the planet's eccentricity vector due to an external perturber is given by \citet{Heggie1996}:
\begin{equation}
    \dot{\mathbf{e}} = \frac{m_3}{M_{12} R^3} \left[6\frac{(\mathbf{r}\cdot\mathbf{R})(\dot{\mathbf{r}}\cdot\mathbf{R})}{R^2}\mathbf{r} - 3\frac{(\mathbf{r}\cdot\mathbf{R})^2}{R^2}\dot{\mathbf{r}} + r^2\dot{\mathbf{r}}\right].
\end{equation}

Following the same procedure used for calculating the energy kick, we find that the change in the planet's eccentricity from a single encounter is
\begin{equation}
    \delta{e} = 3\pi\frac{m_3 a^3}{M_{12}a'^3}
     \left[\frac{(6+9e'^2)(n/n')}{e'^2(1-e'^2)^{3/2}} + \frac{5(n/n')^2}{e'^2}\right]  \Exp{-\frac{n}{n'}\left(\cosh^{-1}(1/e')-\sqrt{1-e'^2}\right)}.
\end{equation}
Substituting this result into Equation~\eqref{eq:delta_a_max}, we obtain the maximum fractional change in semimajor axis for a planet initially on a circular orbit, coplanar with the perturber:
\begin{equation}
\begin{aligned}
    \frac{\delta a_{\max}}{a} = &~ 
    \frac{90\pi^2}{e'^4} \left(\frac{m_3}{M_{123}}\right)^2 \left[1+\frac{n'}{n}\frac{e'^2}{3(1-e'^2)^{3/2}}\right]
    \left[1+\frac{n'}{n}\frac{6+9e'^2}{5(1-e'^2)^{3/2}} \right] \\
    &~ \times \Exp{-2\frac{n}{n'}\left(\cosh^{-1}(1/e')-\sqrt{1-e'^2}\right)}.
\end{aligned}
\end{equation}

\section{Contour Integration}\label{sec:contourintegrals}

This section is dedicated to the derivation of Equation~\eqref{eq:res}.
While the calculation might appear to be a straightforward application of complex analysis, we encountered several subtleties in the derivation that warrant detailed discussion. We document them here for interested readers.

The typical strategy for evaluating integrals of the form in Equation~\eqref{eq:app_deltaeps_termform} is the method of steepest descent, where one expands the integrand about its saddle point $E'_0$ as was done in \citet{Heggie1996, Heggie2006} for hyperbolic encounters.
For the elliptic perturbers considered here, however, the standard Gaussian integral approximations used near the saddle point are inadequate. The difficulty arises from the strongly oscillatory character of the integrand, introduced by the sinusoidal functions $f_i(E)$ in the numerator.
Moreover, including higher-order terms in the phase expansion or retaining additional contributions in the algebraic manipulation of the integrand (such as more terms when integrating by parts to reduce the order of the singularity at $R=0$) do not resolve the discrepancy.
These challenges indicate that a different strategy is required. In what follows, we develop a much simpler approach to solve Equation~\eqref{eq:app_deltaeps_termform}.

\begin{figure}
    \centering
    \begin{tikzpicture}[scale=0.5]
        \draw[thick, <->, gray] (-5, 0) -- (5, 0);
        \draw[thick, <->, gray] (0, -1) -- (0, 8);
        \node[right] at (5, 0) {\Large $\Re E'$};
        \node[above] at (0, 8) {\Large $\Im E'$};
        \filldraw (3, 0) circle(0.2);
        \filldraw (-3, 0) circle(0.2);
        \filldraw (3, 7) circle(0.2);
        \filldraw (-3, 7) circle(0.2);
        \draw[ultra thick, ->] (-3, 0) -- (1.5, 0);
        \draw[ultra thick] (1.5, 0) -- (3, 0);
        \draw[ultra thick, ->] (3, 0) -- (3, 4);
        \draw[ultra thick] (3, 4) -- (3, 5);
        \draw[ultra thick, dashed] (3, 5) -- (3, 7);
        \draw[ultra thick, dashed, ->] (3, 7) -- (-1.5, 7);
        \draw[ultra thick, dashed] (-1.5, 7) -- (-3, 7);
        \draw[ultra thick, dashed] (-3, 7) -- (-3, 5);
        \draw[ultra thick, ->] (-3, 5) -- (-3, 4);
        \draw[ultra thick] (-3, 4) -- (-3, 0);
        \node[below] at (3, -0.13) {\Large $\pi$}; 
        \node[below] at (-3, 0) {\Large $-\pi$};
        \node[above right] at (3, 7) {\Large $\pi + iY$};
        \node[above left] at (-3, 7) {\Large $-\pi + iY$};
        \node[below] at (1, 0) {\Large $\mathcal{C}_1$};
        \node[right] at (3, 2) {\Large $\mathcal{C}_2$};
        \node[above] at (1.5, 7) {\Large $\mathcal{C}_3$};
        \node[left] at (-3, 2) {\Large $\mathcal{C}_4$};
        \filldraw[red] (0, 2) circle(0.2);
        \node[above right, red] at (0, 2) {\Large $E'_0$};
    \end{tikzpicture}
    \caption{
    The four legs $\mathcal{C}_i$ constituting the contour $\mathcal{C}$ that is used to evaluate the integral given by Equation~\eqref{eq:app2_simplified_integral}, where the directions of the contours are indicated by arrows.
    The dashed lines are used to indicate that the limit $Y \to \infty$ is to be taken.
    The pole of the integrand $\mathcal{I}$ at $E'_0$ is indicated by the red dot.
    }\label{fig:contour}
\end{figure}
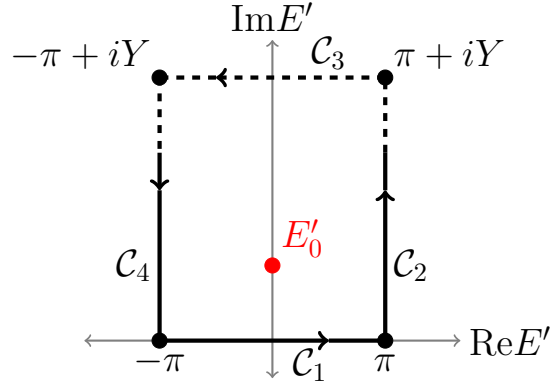

We will demonstrate our approach for the simplest case where $f(E') = 1$, such that
\begin{align}
    I
        &=
            \Int \frac{1}{R^k}\Exp{\F}\,dE' \equiv \frac{1}{(a')^k}\Int \mathcal{I}\,dE',\label{eq:app2_simplified_integral}\\
    \mathcal{I}
        &=
            \frac{\Exp{\F}}{(1 - e' \cos E')^k}.
\end{align}
The key is to evaluate the contour integral (whose value is known by the Cauchy residue theorem)
\begin{align}
    \oint\limits_\mathcal{C} \mathcal{I}\,dE' = 2\pi i\,{\rm Res}[\mathcal{I}; E'_0],\label{eq:oint_contour}
\end{align}
where the contour $\mathcal{C}$ consists of the four legs $\mathcal{C}_1$, $\mathcal{C}_2$, $\mathcal{C}_3$, and $\mathcal{C}_4$ as shown in Figure~\ref{fig:contour} (we will eventually take $Y \to \infty$), and $E'_0$ is the single pole of $\mathcal{I}$:
\begin{equation}
    E'_0 = i\cosh^{-1}\frac{1}{e'}.
\end{equation}

The remaining task is then to evaluate Equation~\eqref{eq:oint_contour} along the three newly-introduced legs of the contour $\mathcal{C}$.
To do so, we introduce the modified integrand 
\begin{align}
    \mathcal{I}'
        &\equiv
            \frac{1}{(-e' \exp\{-iE'\} / 2)^k}
            \Exp{i\frac{n}{n'} \left(E' - e' \frac{\exp\{-iE'\}}{2i}\right)}.
\end{align}
Note that $\mathcal{I}'$ is entire (i.e., free of poles), and so its contour integral along $\mathcal{C}$ must vanish, while its integral along $\mathcal{C}_3$ equals that of $\mathcal{I}$ as $Y \to \infty$ by construction.
Sparing some straightforward algebra, this is sufficient to show that the integrals of $\mathcal{I}$ along $\mathcal{C}_2$, $\mathcal{C}_3$, and $\mathcal{C}_4$ all vanish, and so
\begin{align}
    \int\limits_{\mathcal{C}_1} \frac{1}{R^k}\Exp{\F}\,dE'
    \approx 2\pi i\,{\rm Res}\left[\frac{1}{R^k}\Exp{\F}; {E'_0}\right]
\end{align}
To generalize this result to arbitrary $f(E')$, as stated in Equation~\eqref{eq:res}, we note that any function $f(E')$ that is entire and grows more slowly than $\exp(\exp y)$ does not alter the calculation above. 
As such, it can be directly added to the preceding analysis, giving Equation~\eqref{eq:res}.
Numerical quadrature shows that Equation~\eqref{eq:res} is generally accurate to within a few percent and worsens to no more than a factor of 2 at the very edges of the parameter space.

\bibliography{Bib}
\bibliographystyle{aasjournal}

\end{CJK*}
\end{document}